\documentclass[12pt,a4paper]{article}

\usepackage[T1]{fontenc}
\usepackage{lmodern}
\usepackage{amsmath,amsthm,amssymb,mathtools}
\usepackage[a4paper,margin=2.5cm]{geometry}
\usepackage{setspace}
\usepackage[round,authoryear,longnamesfirst]{natbib}
\usepackage{booktabs,tabularx,longtable,array}
\usepackage{graphicx}
\usepackage{microtype}
\usepackage{enumitem}
\usepackage{placeins}
\usepackage[colorlinks=true,linkcolor=blue,citecolor=blue,urlcolor=blue]{hyperref}
\usepackage{xurl}
\hypersetup{
  pdftitle={When an effort-maximizing team architecture is unstable},
  pdfauthor={Siwei Chen and Pengbo Wang},
  pdfsubject={Contest architecture and team stability},
  pdfkeywords={contest architecture, endogenous teams, coalition formation, team stability, organizational design}
}

\renewcommand{\bibfont}{\normalsize}
\setlist{nosep,leftmargin=*}
\numberwithin{equation}{section}
\allowdisplaybreaks
\newtheorem{theorem}{Theorem}
\newtheorem{proposition}{Proposition}
\newtheorem{corollary}{Corollary}
\newtheorem{lemma}{Lemma}
\theoremstyle{definition}
\newtheorem{definition}{Definition}
\theoremstyle{remark}

\newcommand{\qB}{q_B}
\newcommand{\qG}{q_G}
\newcommand{\qBt}{q_B^{(2)}}
\newcommand{\qGt}{q_G^{(2)}}
\newcommand{\target}{\pi_n}

\newcommand{\ones}{\mathbf{1}}

\begin{document}
\pagenumbering{roman}
\title{When an effort-maximizing team\\architecture is unstable}
\author{Siwei Chen\thanks{Corresponding author. Lingnan College, Sun Yat-sen University, Guangzhou 510275, China. Email: \href{mailto:chensw6@mail.sysu.edu.cn}{chensw6@mail.sysu.edu.cn}. ORCID: \href{https://orcid.org/0000-0002-8486-2251}{0000-0002-8486-2251}.} \and Pengbo Wang\thanks{Lingnan College, Sun Yat-sen University, Guangzhou 510275, China. Email: \href{mailto:illusory0101@gmail.com}{illusory0101@gmail.com}.}}
\date{October 2026}
\maketitle\thispagestyle{empty}
\begin{abstract}
\normalsize\noindent
Organizations choose team structures to intensify competition, yet their
members may prefer to reorganize. We ask whether an architecture that
maximizes aggregate effort can remain stable when contestants may form a
replacement team from incumbent rosters. When a
coalition draws from identical teams and reproduces their type, its
members and the designer rank the reorganization in opposite directions:
the coalition gains precisely when total effort falls. A strict effort optimum
can therefore supply its own blocking certificate, while a stable optimum can
survive only on a plateau where every reachable reorganization of this kind
leaves effort unchanged. In a linear benchmark, we
characterize stability against every feasible one-team reorganization and
global effort maximization over every feasible partition. Except under pair
capacity, the two regions do not overlap; with pair capacity, they meet only at
one boundary. The ordering persists in a matched quadratic benchmark with
strictly convex effort cost, and broader one-step deviation rights cannot
restore overlap. Among standardized full-cap designs, the unique stable-effort
optimum uses the largest feasible roster and sets incentive concentration at
its blocking boundary.

\end{abstract}
\clearpage
\pagenumbering{arabic}

\section{Introduction}\label{sec:introduction}

An effort-maximizing team architecture can give its own members a reason to
dismantle it.  A designer values total competitive effort, whereas contestants
value their own continuation payoffs.  When a coalition can rebuild an
incumbent team without changing its prize or incentive technology, it may gain
by weakening the competitive environment around it.  We show that these two
objectives rank the same reorganization in opposite directions.  A strict
effort optimum can therefore supply its own blocking certificate.

Consider an institution with a fixed population, roster capacity, governance
rule, and formation rights.  Within those rules, a designer selects a target
partition, while a coalition may recruit across incumbent teams and form one
new team.  A capacity rule can prohibit a larger team and still allow a
replacement of exactly the permitted size.  For example, a four-person clone
can take one contestant from one full roster and three from another, leaving
residual teams of sizes three and one.  Its own prize and incentives are
unchanged, but two full rivals become smaller residual teams.  Because the
active role is assigned after formation, members of identical source teams
share the same ex ante payoff when they reorganize.

For clone deviations from identical source teams, every deviator begins with
the same payoff and joins a team whose payoff falls with aggregate effort.
The coalition therefore blocks if and only if the reorganization lowers
effort.  A stable effort optimum can survive only on a plateau, while a strict
optimum with a feasible clone deviation is unstable.  In the homogeneous
size-dependent model, two identical non-singleton teams supply a simple block:
one source contributes one member and another contributes the rest of the new
team.

The linear benchmark fully characterizes both comparison sets.  Team prizes
scale with size, effort costs are linear, and the governance rule shifts the
winning prize toward the member who acts for the team---from equal sharing to
full incentive concentration.  We derive one exact boundary for stability
against every capacity-feasible coalition forming one new team and another
for global aggregate-effort optimality among every capacity-feasible
partition.  The coalition boundary is set by a concentrated two-source
replacement, whereas the designer boundary is set by a near-balanced partition
with one additional team.  Except under pair capacity, the first boundary lies
strictly below the second, so stability and even weak global optimality do not
overlap.  Under pair capacity the weak sets meet only at their common boundary.
Expanding the admissible deviation class cannot restore the non-overlap.  A
matched quadratic specification reproduces the threshold ordering with strictly
convex effort cost.

Capacity regulates team size, whereas mobility and contracting rules determine
where a replacement may recruit.  Restricting recruitment to one incumbent
protects a full-cap target from a nontrivial full-size clone; allowing a second
source makes the canonical replacement feasible without changing the contest
technology.

When roster scale and governance are chosen together, the stable standardized
design with the largest feasible roster---and hence the fewest equal
teams---generates the most effort.  The Online Appendix develops
matched-curvature extensions and exact results for source-span restrictions and
retention frictions.

\subsection{Related literature}\label{subsec:literature}

Prize sharing and contest architecture shape rent-seeking effort
\citep{nitzan1991sharing,moldovanu2006contest}.  \citet{deng2025monopoly} ask
whether contests that maximize effort for a monopolistic designer remain
viable when contest designers compete for participants.  Like that paper, we
open an organizational margin around an effort-maximizing design, but here it
comes from voluntary re-formation within a fixed population rather than
competition among contest designers.  The continuation analysis uses the
replacement- and share-function methods for asymmetric contests
\citep{szidarovszky1997existence,cornes2005asymmetric,
yamazaki2008asymmetric}.

Endogenous group formation can change aggregate dissipation.
\citet{baik1995competitive} obtain a voluntarily formed competitive-share group
with lower dissipation than individual competition.  \citet{baik1997collective}
endogenize group size through mobility in a two-group contest;
\citet{baik2001strategic} endogenize the number and sizes of strategic groups and
show that formation can raise or lower dissipation.  \citet{chowdhury2017small} show how splitting contestants
and prizes across separate sub-contests can raise total effort.  Here equal
full-cap teams compete for a common prize and must withstand coalition replacement:
equal sharing favors smaller rosters, but the ranking reverses along their
stability frontiers.
\citet{sanchezpages2007endogenous} studies which coalition structures survive
under alternative defection and residual-response conventions.  That literature
asks which groups form and how formation changes conflict or dissipation.  Our
analysis starts from a designer-selected architecture and shows that a type-preserving
cross-source clone ranks continuations exactly opposite to the designer's
effort objective.  The linear benchmark then compares every feasible block
with every feasible architecture.  Related work studies stable group formation
and internal conflict \citep{garfinkel2004stability,bloch2006peace,
morelli2016internal,lee2022group}.  Competition between teams and alliance
formation provide complementary models \citep{eliaz2018simple,
chade2020competing,konishi2020sequential,konishi2021endogenous,dollinger2026alliances,
sun2026alliance}.

The closest recent subject-matter neighbor, \citet{konishi2026formation}, fixes
team capacity and a common sharing rule and studies ability sorting under
head-hunting and swaps.  Here homogeneous contestants recruit across incumbent
rosters to form a replacement team while the residual teams remain in the
contest.  The resulting conflict is between the designer's effort objective
and the members' common payoff ranking rather than between sorting patterns.
Coalition-structure models provide the broader externality language
\citep{bloch1996sequential,yi1997stable,ray1999theory,ray2015coalition}.
Hedonic, matching, and network models supply complementary stability and consent
comparisons \citep{bogomolnaia2002stability,
jackson1996strategic,dutta1997stable,andelman2009strong}.
In these comparison frameworks, outcomes are ranked by participant payoffs or
surplus \citep{hafalir2007efficiency,
sakamoto2025selfenforcing,nevrekar2022efficiency}; here the designer ranks
architectures by equilibrium aggregate effort, which the coalition cannot
redistribute.

\section{Model}\label{sec:environment}

\subsection{Teams, payoffs, and timing}\label{subsec:timing}

There are \(N=Mn\) risk-neutral contestants, where \(M,n\geq2\).  A feasible
architecture is a partition \(\pi=(S_1,\ldots,S_J)\) into nonempty teams with
\(|S_j|\leq n\).  The upper bound \(n\) is both the target team size and the
institutional roster capacity.  A size-\(k\) team receives total prize
\(P_k>0\) if it wins.

Before effort is chosen, one member of each team is selected uniformly as its
active role.  The active member chooses the team's effort.  Let
\(a_k\in(0,P_k]\) denote the winning-state value that governs this choice.
The residual prize can be shared among the remaining members.  Because the role
is assigned symmetrically before effort, a member's formation payoff averages
both prize receipts and the active member's effort cost over the \(k\) possible
role realizations.

The lottery captures a team formed before an execution-stage lead is selected,
preserving ex ante symmetry while the governance rule concentrates
performance-linked compensation on that lead.

Our benchmark governance rule is indexed by \(q\in[0,1]\).  The active role
receives the fraction \(q+(1-q)/k\) of the team prize, while every other member
receives \((1-q)/k\).  Hence
\begin{equation}\label{eq:effective-value-general}
  a_k=P_k\left(q+\frac{1-q}{k}\right).
\end{equation}
Governance changes the contest value \(a_k\) without changing the total prize
\(P_k\).  At \(q=0\), the prize is shared equally; at \(q=1\), the active
member receives the entire winning prize.  Thus \(q\) measures incentive
concentration within the team.  The ranking reversal holds beyond this
parameterization; the exact benchmarks use it to locate the two institutional
thresholds.

The sequence of play is summarized below.
\begin{center}
\renewcommand{\arraystretch}{1.15}
\begin{tabularx}{\textwidth}{@{}p{.09\textwidth}p{.21\textwidth}X@{}}
\toprule
Time & Decision maker & Decision or event \\
\midrule
0 & Institutional environment
& Fixes the population, roster capacity, governance rule, and permitted
formation moves. \\
1 & Designer
& Chooses a target partition from the capacity-feasible design class. \\
2 & Contestants
& May implement one permitted formation move, thereby determining the team
architecture for the contest. \\
3 & Nature
& Draws one active member uniformly from each team; roles and winning-state
shares are observed. \\
4 & Active members
& Choose team efforts simultaneously; winning probabilities, prizes, and
effort costs are realized. \\
\bottomrule
\end{tabularx}
\end{center}
Any formation move initiates a new continuation contest.  Nonempty residual
teams remain in the game, and every team's effort and winning probability are
recomputed under the resulting architecture.  This recomputation is the
externality that gives a split--clone deviation its force.
The baseline design problem holds this institutional environment fixed and
asks which target partition maximizes aggregate effort and whether contestants
preserve it.  Later comparisons vary one institutional margin at a time.

\subsection{The continuation contest}\label{subsec:continuation}

For the general mechanism result, values and costs may differ across teams;
the exact benchmarks below specialize to a common cost.  Every effective value
satisfies \(a_j>0\).  Each cost \(c_j\) is continuous on \([0,\infty)\), is
\(C^2\) on \((0,\infty)\), and is normalized by \(c_j(0)=0\), with
\(c_j'(x)>0\) and \(c_j''(x)\geq0\) for \(x>0\).  These are the standard smooth
convex-cost conditions for an asymmetric contest.  For the general interior
continuation result, we additionally impose
\(c'_{j,+}(0)=0\).  This zero-origin marginal-cost condition selects the
familiar no-dropout case in which every team is active.

For an architecture \(\pi\), team \(j\)'s active member chooses effort
\(x_j\geq0\).  Write \(X=\sum_hx_h\) and \(p_j=x_j/X\) when \(X>0\).
At an all-zero profile, a unilateral positive effort wins for sure; assigning
equal winning probabilities at that profile gives the same equilibrium
conclusions.
The active member solves
\begin{equation}\label{eq:contest-problem}
  \max_{x_j\geq0}\;a_j\frac{x_j}{x_j+X_{-j}}-c_j(x_j).
\end{equation}
Conditional on an architecture, this is exactly a finite asymmetric Tullock
contest with team-specific values \(a_j\) and costs \(c_j\).  Team formation
determines which contestants share each continuation payoff and which values
enter that contest.
Whenever the continuation equilibrium is interior, its first-order condition is
\begin{equation}\label{eq:general-foc}
  a_j(1-p_j)=Xc_j'(p_jX).
\end{equation}
The ex ante payoff of any member of team \(j\) is
\begin{equation}\label{eq:member-payoff}
  u_j=\frac{P_jp_j-c_j(p_jX)}{k_j},
  \qquad k_j=|S_j|,
\end{equation}
where \(P_j=P_{k_j}\) in the size-dependent model.

Equation~\eqref{eq:member-payoff} separates the active role's incentive value
from the team prize: effort responds to \(a_j\), while formation payoffs
distribute \(P_j\) and average effort cost across members ex ante.

\begin{lemma}[Unique interior continuation]\label{lem:continuation}
Under the cost regularity and zero-origin condition above, the exponent-one
Tullock continuation game has a unique
pure-strategy Nash equilibrium, and every team chooses strictly positive
effort.  At that equilibrium, \(X>0\) is single-valued and
\[
 a_j(1-p_j)=Xc_j'(p_jX),\qquad p_j\in(0,1),\qquad
 \sum_{j=1}^Jp_j=1.
\]
\end{lemma}

Lemma~\ref{lem:continuation} restates the interior asymmetric-contest results of
\citet[Proposition~1]{yamazaki2008asymmetric} and
\citet[Proposition~3(5)]{cornes2005asymmetric} in the present notation.
Appendix~\ref{ec:continuation} gives the coordinate mapping and a direct
verification.

The zero-origin condition is used only to obtain this single-valued interior
map.  The linear benchmark instead allows inactive teams and is handled by the
active-set formula below.

For unit cost, \(c(x)=x\), a pure-equilibrium active set \(\mathcal A\) with
\(m\geq2\) satisfies
\begin{equation}\label{eq:linear-continuation}
  X=\frac{m-1}{\sum_{h\in\mathcal A}1/a_{|S_h|}},
  \qquad
  p_j=1-\frac{X}{a_{|S_j|}},
  \qquad
  x_j=Xp_j.
\end{equation}
An active team has \(a_{|S_j|}>X\); equality is assigned to the inactive side.
For quadratic cost, \(c(x)=x^2/2\), the unique interior intensity
\(\lambda=X^2\) solves
\begin{equation}\label{eq:quadratic-continuation}
  \sum_{j=1}^{J}\frac{a_{|S_j|}}{a_{|S_j|}+\lambda}=1,
  \qquad
  p_j=\frac{a_{|S_j|}}{a_{|S_j|}+\lambda}.
\end{equation}

Linear cost may leave some teams inactive, whereas quadratic cost yields an
interior solution.  In both cases, a formation move is evaluated in the
continuation generated by the new architecture.

\subsection{Formation moves and design criteria}\label{subsec:formation}

Formation rights determine which partition changes contestants can implement.
The paper allows one coalition to form one new team by recruiting from several
incumbents, with every nonempty residual team remaining in the recomputed
continuation.  This cross-source move captures coordinated replacement while
preserving a one-step formation problem.

The initial target is
\[
  \target=(\underbrace{n,\ldots,n}_{M\text{ teams}}).
\]
A \emph{one-new-team deviation} selects a coalition from one or more incumbent
teams and forms one new team of size at most \(n\).  Every incumbent team keeps
its nondeviating members if any remain, and the entire continuation contest is
recomputed.  A deviation is a strict block if every coalition member obtains a
strictly higher ex ante payoff.

Represent a deviation by its extraction vector.  If the coalition
takes \(t_i\geq1\) members from source \(i=1,\ldots,\ell\), its new team has
size \(s=\sum_{i=1}^{\ell}t_i\leq n\), while source \(i\) leaves a residual of
size \(n-t_i\) whenever \(t_i<n\).  The number \(\ell\) is the deviation's
\emph{source span}.  Extraction vectors that differ only by source labels are
identified.  A full-size replacement has \(s=n\); it is nontrivial only when it
uses at least two sources.

\begin{definition}[One-new-team stability]\label{def:stability}
The target \(\target\) is one-new-team stable if no capacity-feasible coalition
can form one new team and strictly benefit all its members in the recomputed
continuation.
\end{definition}

The residual-response convention is \(\delta\)-type: contestants left behind
remain with their residual incumbent teams while the continuation contest is
recomputed \citep{hart1983endogenous,ray2015coalition}.  Relative to the
internal blocking analyzed by \citet{ray2015coalition}, a split--clone
coalition recruits from multiple incumbents and generally produces neither a
refinement nor a coarsening of the original partition.  Related
bounded-deviation traditions appear in
\citet{dreze1980hedonic,andelman2009strong}.

\begin{definition}[Global effort optimality]\label{def:global-optimality}
Let \(\Pi_{Mn,n}\) denote all partitions of \(Mn\) contestants into nonempty
teams of size at most \(n\).  The target is a global aggregate-effort maximizer
if \(X(\target)\geq X(\pi)\) for every \(\pi\in\Pi_{Mn,n}\).  It is a strict
global maximizer modulo labels if equality implies that \(\pi\) has the same
multiset of team sizes as \(\target\).
\end{definition}

At fixed capacity, governance, and formation rules, the designer and
contestants face different comparison sets.  The designer compares the target
with every feasible partition; contestants compare it with the continuations
they can reach under the formation technology.  The central question is
whether these two rankings can select the target at the same parameter values.
Aggregate effort is a natural objective when competitive activity produces
innovation, customer acquisition, or campaign mobilization.

\section{Clone deviations and the ranking reversal}\label{sec:splitclone}

A designer maximizing aggregate effort and a coalition able to clone an
incumbent team rank the same reachable architectures in opposite directions.
A strictly effort-maximizing architecture therefore supplies its own blocking
certificate.  The result applies when the deviators come from incumbent teams
of one common type and reproduce that type.

A team type is a primitive tuple
\(\tau=(k_\tau,P_\tau,a_\tau,c_\tau)\).  If an architecture \(\pi\)
contains at least two teams of type \(\tau\), let \(\mathcal C_\tau(\pi)\)
be the set of capacity-feasible continuations obtained when a coalition draws
members from at least two such teams, forms one new team of the same type, and
leaves every nonempty residual team in place.  Thus the new team changes its
competitive field while holding its own prize technology fixed.
The formation environment assigns primitives to every residual roster, so
each element of \(\mathcal C_\tau(\pi)\) includes the primitives needed to define
its continuation; in the homogeneous size-dependent benchmarks, a size-\(k\)
residual simply inherits \((P_k,a_k,c_k)\).

\begin{lemma}[Clone payoff reverses the effort ranking]\label{lem:clone}
Under the conditions of Lemma~\ref{lem:continuation}, fix a team type \(\tau\) satisfying
\(P_\tau\geq a_\tau>0\).  Its equilibrium winning probability and each
member's continuation payoff
\[
 u_\tau(X)=\frac{P_\tau p_\tau(X)-c_\tau(p_\tau(X)X)}{k_\tau}
\]
are strictly decreasing in aggregate effort \(X\).
\end{lemma}

As aggregate effort rises, the clone must reduce its equilibrium share while
paying more to sustain any given share.  The condition
\(P_\tau\geq a_\tau\) makes both effects weakly reduce member payoff, with the
cost effect strict.  It holds automatically under the sharing rule in
\eqref{eq:effective-value-general}.  Appendix~\ref{ec:splitclone} gives the
derivative.

\begin{theorem}[Clone--effort ranking reversal]\label{thm:splitclone}
Suppose architectures \(\pi\) and
\(\pi'\in\mathcal C_\tau(\pi)\) induce continuations satisfying
Lemma~\ref{lem:continuation} and
\(P_\tau\geq a_\tau\).  Then the type-identical-source clone deviation to
\(\pi'\) strictly benefits every coalition member if and only if
\[
                 X(\pi')<X(\pi).
\]
\end{theorem}

Teams of one primitive type give their members the same continuation payoff.
Every source member therefore compares \(u_\tau(X(\pi))\) with the clone payoff
\(u_\tau(X(\pi'))\), and Lemma~\ref{lem:clone} gives the equivalence.

\begin{corollary}[Effort optimality and stability]
\label{cor:plateau}
Let \(\Pi\) be a feasible design class, and suppose \(\pi\in\Pi\) maximizes
\(X\) over \(\Pi\).  If \(\pi\) is stable against every reachable
type-identical-source clone split, then
\[
 X(\pi')=X(\pi)
 \quad\text{for every }\tau\text{ and every }
 \pi'\in\mathcal C_\tau(\pi)\cap\Pi.
\]
If the maximizer is strict and
\(\mathcal C_\tau(\pi)\cap\Pi\neq\varnothing\) for some \(\tau\),
then \(\pi\) is unstable.  If \(\Pi\) contains all capacity-feasible partitions,
then in the homogeneous size-dependent model any strict effort maximizer
containing two identical non-singleton teams is blocked by the canonical split
that takes one member from one of those teams and the remaining \(k-1\)
members from the other.
\end{corollary}

Optimality places every reachable clone weakly below the original effort, while stability rules
out a strict decrease.  A stable optimum must therefore lie on a plateau.  At
a strict optimum, any feasible clone deviation changes the team-size multiset,
lowers effort, and blocks.  Appendix~\ref{ec:splitclone} gives the formal
distinctness argument and the canonical homogeneous instance.

The common-source condition makes the coalition's comparison scalar and is
tight, as the capacity-feasible heterogeneous-source example in
Appendix~\ref{ec:splitclone} shows.

The clone keeps its prize rule and uses no transfers, so its profitability
comes from the continuation externality imposed on residual opponents.  This
turns the coalition-structure externality emphasized by
\citet{ray2015coalition} into a sharp implication for a designer who ranks
architectures by aggregate effort rather than by welfare or transferable value
\citep{dutta1997stable,bogomolnaia2002stability,jackson1996strategic,
andelman2009strong}.

\section{Exact thresholds and non-overlap}\label{sec:linear-stability}

The ranking reversal establishes the conflict in general.  The benchmark
analysis now identifies the coalition that blocks first and the architecture
that is the designer's strongest rival.

\subsection{Linear stability threshold}

Set
\begin{equation}\label{eq:linear-benchmark}
  P_k=k,\qquad a_k=1+q(k-1),\qquad c(x)=x,
  \qquad q\in[0,1].
\end{equation}
The prize scaling treats a size-\(k\) team as commanding \(k\) times the
opportunity of a singleton---for example, because roster size scales the
project or market opportunity at stake.  Combined with the sharing rule, it
makes the active-role value affine in team size.
Although unit cost lies outside the zero-origin condition in
Lemma~\ref{lem:continuation}, its active-set formula supplies the same local
ranking reversal.  Whenever a source type and its clone are active,
\[
 p_k=1-\frac{X}{a_k},
 \qquad
 U_k(p)=p\{1-r_k(1-p)\}
\]
make the clone's probability strictly decreasing in \(X\) and its member
payoff strictly increasing in \(p\).  The exact envelope below separately
handles inactive residual teams and the complete one-new-team deviation class.
The target contains \(M\) identical value-\(a_n\) teams.  A general deviation
draws \(t_i\geq1\) members from each of \(\ell\) incumbent sources and forms one
team of size \(s=\sum_i t_i\leq n\).  Nonempty residual teams have sizes
\(n-t_i\).  This description includes one-source splits, smaller cross-source
teams, and full-size replacements.

The target continuation is symmetric:
\begin{equation}\label{eq:linear-target}
  p_n^0=\frac1M,
  \qquad
  X_0=\frac{M-1}{M}a_n,
  \qquad
  x_n^0=\frac{M-1}{M^2}a_n.
\end{equation}
For a full-size replacement, every deviator joins a clone of a target team.
Thus its members strictly gain exactly when the replacement's probability
exceeds \(1/M\).  At the target effort level this is equivalent to the
residual teams contributing too little probability mass to sustain the
original intensity.  The coalition calculation can therefore be written as a
sum of residual losses rather than as \(n\) separate member comparisons.

The first potential block is canonical: one member leaves one incumbent and
\(n-1\) members leave another.  The coalition forms a new size-\(n\) team and
leaves residual teams of sizes \(n-1\) and \(1\).  Its blocking margin changes
sign at the positive root of
\begin{equation}\label{eq:qB-polynomial}
  (M-1)(n-1)(n-2)q^2+[n(M-2)+2]q-1=0.
\end{equation}
Define
\begin{equation}\label{eq:qB-def}
  \qB=
  \frac{2}{
  \sqrt{[n(M-2)+2]^2+4(M-1)(n-1)(n-2)}
  +n(M-2)+2}.
\end{equation}
This expression also covers \(n=2\), when the quadratic coefficient vanishes.

\begin{theorem}[Exact linear stability threshold]\label{thm:linear-stability}
Under \eqref{eq:linear-benchmark}, the target \(\target\) is stable against all
capacity-feasible one-new-team deviations if and only if \(q\leq\qB\).
At \(q=\qB\), the canonical \(1+(n-1)\) full-size replacement is indifferent
and no deviation is strictly profitable.  For \(q>\qB\), that coalition
strictly blocks.  No one-source split or smaller cross-source deviation starts
blocking earlier.
\end{theorem}

The canonical split preserves the new team's full size while concentrating
the damage to the teams left behind.  One source is reduced to a singleton and
the other supplies the remaining members, producing the weakest residual
field available to a nontrivial full-size replacement.  Appendix
\ref{ec:linear-stability} proves that no smaller team, one-source move, or
active-set change crosses earlier.

Source span changes the result discontinuously.  If one new team may recruit
from only one incumbent source, the full-cap target
is protected from nontrivial replacement.  Once recruitment from two sources
is permitted, the exact onset of instability is \(\qB\).

The threshold is the positive root of \(Aq^2+Bq-1=0\), with
\[
 A=(M-1)(n-1)(n-2),
 \qquad
 B=n(M-2)+2.
\]
Increasing either the number of incumbent teams or capacity weakly raises both
coefficients and strictly raises at least one, lowering \(\qB\).  Larger fields
and rosters thus make cross-source replacement profitable at
lower incentive concentration: the clone stays at capacity, but extraction
creates a larger change in the effective opposition.

At \(q=0\), every nonempty team has the same active-role value.  Splitting
incumbents then adds competitors without selectively weakening their values,
so the target is stable.  At \(q=1\), effective value is proportional to team
size; a small residual loses most of its competitive force, and the canonical
replacement is profitable.  The unique crossing between these endpoints is
\(\qB\).  For \(n=2\), the canonical extraction is necessarily balanced,
which is why several strict comparisons in later sections collapse to
equality.

\subsection{Global optimality and threshold separation}
\label{sec:linear-optimality}

The designer may compare \(\target\) with every partition of \(Mn\) contestants
into teams of size at most \(n\).  For a fixed number of active teams, the
linear continuation in \eqref{eq:linear-continuation} makes effort depend on
the harmonic sum of effective values.  Since \(1/a_k\) is strictly convex in
team size under \eqref{eq:linear-benchmark}, the best partition with a fixed
number of teams is as balanced as integer feasibility permits.  A perspective
convexity argument then shows that the first global competitor has exactly
\(M+1\) active teams.

To make the reduction explicit, let a candidate have \(J\geq2\) active teams
with values \(a_{k_1},\ldots,a_{k_J}\).  Its aggregate effort is
\begin{equation}\label{eq:linear-partition-effort}
 X(k_1,\ldots,k_J)
 =
 \frac{J-1}{\sum_{j=1}^{J}1/a_{k_j}}.
\end{equation}
For fixed \(J\) and \(\sum_jk_j=Mn\), maximizing effort is equivalent to
minimizing the denominator.  Since
\[
 k\longmapsto\frac{1}{1+q(k-1)}
\]
is strictly convex for \(q>0\), any pair of team sizes differing by at least
two can be moved one unit toward each other and strictly lower the denominator.
The maximizing integer partition must therefore use only the two adjacent
sizes \(\lfloor Mn/J\rfloor\) and \(\lceil Mn/J\rceil\).  At \(q=0\), all team
values coincide and the team count alone determines effort; the same balanced
representative can be selected without loss.

Any partition different from the full-cap target has at least \(M+1\) teams.
Among the balanced candidates, the \(M+1\)-team partition is the last one the
target overtakes as incentive concentration rises.  It therefore sets the
global boundary.  Appendix~\ref{ec:linear-global} proves the comparison for
every feasible team count and integer remainder.

Let
\begin{equation}\label{eq:linear-balanced-competitor}
  d=\left\lfloor\frac{Mn}{M+1}\right\rfloor,
  \qquad
  r=Mn-(M+1)d.
\end{equation}
The first competitor contains \(M+1-r\) teams of size \(d\) and \(r\) teams of
size \(d+1\).  Write \(a_k(q)=1+q(k-1)\) and define
\begin{equation}\label{eq:G-polynomial}
\begin{split}
  G_{M,n}(q)
  ={}&M^2a_d(q)a_{d+1}(q)\\
     &-(M-1)a_n(q)
       \big[(M+1-r)a_{d+1}(q)+r a_d(q)\big].
\end{split}
\end{equation}
Appendix~\ref{ec:linear-global} expands this quadratic, proves that it has a
unique root in \((0,1)\), and gives the closed form.  Denote that root by
\(\qG\).

\begin{theorem}[Exact linear global threshold]\label{thm:linear-global}
Under \eqref{eq:linear-benchmark}, the target \(\target\) is a global
aggregate-effort maximizer among all capacity-feasible partitions if and only
if \(q\geq\qG\).  At \(q=\qG\), it ties the near-balanced \(M+1\)-team
partition in \eqref{eq:linear-balanced-competitor}.  For \(q>\qG\), it is the
unique maximizer modulo team and member labels.
\end{theorem}

Different architectures set the two boundaries.  A coalition gains by
concentrating departures on two incumbents and leaving weak residual teams;
the designer's threshold is set by spreading the population as evenly as
possible over one additional team.  Cross-multiplication gives \(G_{M,n}(q)\).
Its endpoint signs and the integer-remainder identities establish a unique
crossing, with the near-balanced competitor active throughout; Appendix
\ref{ec:linear-global} gives the details.

\begin{theorem}[Exact linear non-overlap]\label{thm:linear-gap}
For all \(M,n\geq2\), \(\qB\leq\qG\).  Equality holds when \(n=2\), and
\[
  \qB<\qG
  \qquad\text{when }n>2.
\]
Hence one-new-team stability never overlaps with \emph{unique} global effort
optimality.  When \(n>2\), stability does not overlap even with weak global
optimality.
\end{theorem}

The weak ordering is already implied by the ranking reversal.  If
\(q_G<q\leq q_B\), the target would be both a strict global maximizer and
stable even though its canonical clone split is feasible, a contradiction.
The exact comparison sharpens this logic.  At \(q_B\), the canonical split
architecture ties the target in effort and has \(M+1\) teams.  Among
partitions with that team count, convexity makes the near-balanced partition
weakly better, and strictly better unless \(n=2\).  Hence the target has not
reached its global boundary at \(q_B\) when \(n>2\); the unique crossing then
gives \(q_B<q_G\).  Under pair capacity the canonical size vector is itself
near-balanced, producing equality.  Appendix~\ref{ec:linear-global} gives the
complete convexity argument.

\begin{figure}
\centering
\IfFileExists{phase_diagram.pdf}{
  \includegraphics[width=.92\textwidth]{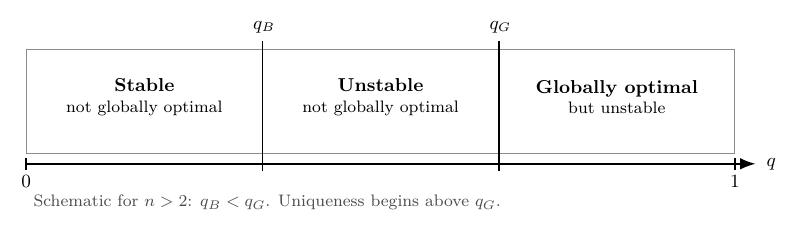}
}{
  \fbox{\parbox{.86\textwidth}{For \(n>2\), the incentive line is divided at
  \(\qB<\qG\): the target is stable but not globally optimal for
  \(q\leq\qB\), unstable and not globally optimal for
  \(\qB<q<\qG\), and globally optimal but unstable for \(q\geq\qG\).
  Uniqueness begins strictly above \(\qG\).}}
}
\caption{The stability--optimality phase diagram for \(n>2\).  Stability
disappears before global effort optimality arrives.  The matched quadratic
benchmark in Section~4.3 has the same ordering.}
\label{fig:phase}
\end{figure}

Throughout the middle region in Figure~\ref{fig:phase}, the target is already
unstable while a different architecture still generates more effort.
Appendix Figure~\ref{fig:linear-exact-thresholds} plots the exact thresholds
over representative capacity and team-count ranges.

For pair capacity the common boundary is explicit.  Substituting \(n=2\) in
\eqref{eq:qB-polynomial} gives
\begin{equation}\label{eq:pair-cap-boundary}
 q_B(M,2)=q_G(M,2)=\frac{1}{2(M-1)}=\frac{1}{N-2}.
\end{equation}
For each fixed \(M\), weak stability and weak optimality therefore meet at
exactly this point.  The canonical coalition is indifferent there.

\begin{corollary}[Larger deviation classes]\label{cor:deviation-expansion}
Let \(\mathcal D_1\) be the one-new-team deviation class and let
\(\mathcal D^+\supseteq\mathcal D_1\) expand only the permitted one-step deviations,
holding the architecture set, continuation game, payoffs, and effort objective
fixed.  Then
\[
 \operatorname{Stable}(\mathcal D^+)
 \subseteq \operatorname{Stable}(\mathcal D_1)=[0,q_B].
\]
For every \(n\geq2\), stability under \(\mathcal D^+\) cannot overlap unique
global effort optimality.  For \(n>2\), it cannot overlap even weak global
effort optimality.
\end{corollary}

\begin{proof}
Every target stable under \(\mathcal D^+\) is stable under its subset
\(\mathcal D_1\), which gives the set inclusion.  Theorem~\ref{thm:linear-stability}
identifies \(\operatorname{Stable}(\mathcal D_1)=[0,q_B]\); combining this
identity with Theorems~\ref{thm:linear-global} and \ref{thm:linear-gap} gives
the two non-overlap conclusions.
\end{proof}

Every expansion therefore inherits the non-overlap conclusion, while its exact
stability threshold may move below \(q_B\).  At pair capacity, additional
deviations determine whether the boundary tie survives.

\subsection{Matched quadratic benchmark}

The same ordering emerges in a matched benchmark with strictly convex effort
cost.  Consider the matched quadratic specification
\begin{equation}\label{eq:quadratic-benchmark-main}
 c(x)=\frac{x^2}{2},\qquad
 P_k=k^2,\qquad
 a_k=k[1+q(k-1)].
\end{equation}
The matched scaling couples a more convex effort technology with more rapidly
growing project stakes, preserving the homogeneity that makes all partitions
exactly comparable.  It tests whether the threshold ordering persists in a
strictly convex contest while retaining a closed global architecture problem.

\begin{proposition}[Exact matched quadratic thresholds]
\label{prop:quadratic-exact}
Under \eqref{eq:quadratic-benchmark-main}, there are unique thresholds
\(\qBt,\qGt\in(0,1)\).  The target is stable against every
capacity-feasible one-new-team deviation if and only if \(q\leq\qBt\), and it
is a global aggregate-effort maximizer if and only if \(q\geq\qGt\).  Below
\(\qGt\), the near-balanced \(M+1\)-team partition strictly outperforms the
target; above \(\qGt\), the target is the unique maximizer modulo labels.
Moreover,
\[
 \qBt<\qGt\quad\text{for }n>2,
 \qquad
 \qBt=\qGt=\frac{1}{2(M-1)}\quad\text{for }n=2.
\]
\end{proposition}

A concentrated two-source replacement and a near-balanced partition with one
additional team continue to determine the boundaries.
Appendices~\ref{ec:quadratic-stability}--\ref{ec:quadratic-gap} derive
the thresholds, prove the ordering, and characterize the boundary ties.

\FloatBarrier
\section{Stable design and institutional implications}\label{sec:scope}
\label{subsec:stable-scale}

The threshold results also yield a positive design prescription when roster
scale and governance are chosen jointly within the standardized full-cap
family.  Fix a population \(N\) and let
\[
 \mathcal D_N=\{d\in\mathbb N:d\mid N,\ 2\leq d\leq N/2\}
\]
be the admissible full-cap roster sizes.  For each \(d\in\mathcal D_N\), the
target consists of \(N/d\) equal teams, the institution chooses the common
governance rule, and the target must be one-new-team stable under roster capacity
\(d\).  Residual teams created by a deviation need only respect that capacity.
The two target-effort functions are
\[
 X_L(d,q)=\frac{N-d}{N}[1+q(d-1)],
 \qquad
 X_Q(d,q)=\sqrt{(N-d)[1+q(d-1)]}.
\]
Both increase with \(q\), so each roster is moved to its exact blocking boundary.

\begin{proposition}[Stable standardized scale design]
\label{prop:stable-scale}
Suppose \(\mathcal D_N\) is nonempty.  In both the linear and matched quadratic
benchmarks, target effort at the largest stable governance rule is strictly
increasing in \(d\in\mathcal D_N\).  Let \(p\) be the smallest prime divisor of
\(N\).  The unique stable-effort optimum within the standardized full-cap family
therefore uses
\[
 d^S=\frac{N}{p},\qquad M^S=p,
\]
and governance \(q_B(p,N/p)\) in the linear benchmark or
\(q_B^{(2)}(p,N/p)\) in the matched quadratic benchmark.  If \(N\) is even,
the design consists of two equal teams.
\end{proposition}

At equal sharing, smaller rosters create more rivals and generate more effort.
The stability-constrained comparison instead moves every roster to its own
blocking boundary.  Along those boundaries the governance adjustment more than
offsets the loss of rival teams, reversing the equal-sharing ranking.
Appendix~\ref{ec:stable-scale} proves the proposition.

Institutions regulate team size, incentive concentration, and recruitment
scope through different instruments.  Roster capacity often reflects
technological or legal constraints, compensation rules govern incentive
concentration, and mobility or contracting rights determine how widely a new
team can recruit.  Confining recruitment to one incumbent protects the full-cap target
from every permitted split in both exact benchmarks.  Once recruitment from a
second incumbent is allowed, the canonical \(1+(n-1)\) replacement becomes
feasible and determines the onset of instability.  Recruitment rights can
therefore change implementation without changing capacity or the continuation
contest.

Effort optimality and one-new-team stability can therefore point in opposite
directions.  For a reachable
type-identical-source clone split, the designer's effort ranking is exactly
reversed by every deviator's payoff ranking.  A stable effort optimum must
therefore sit on a plateau; whenever a feasible clone remains, a strict optimum
is self-undermining.

In the linear benchmark, a concentrated two-source replacement determines
\(q_B\), whereas a balanced \(M+1\)-team partition determines \(q_G\).  Their
strict ordering above pair capacity eliminates overlap between stability and
weak global effort optimality.  The matched quadratic benchmark reproduces
the ordering with strictly convex cost, and enlarging the deviation class
cannot restore overlap.

\section*{Declarations}

\subsection*{Data and code availability}
The paper is theoretical and uses no empirical data.

\subsection*{CRediT authorship contribution statement}
\textbf{Siwei Chen:} Conceptualization, Methodology, Formal analysis,
Validation, Writing--original draft, Writing--review \& editing, Project
administration, Funding acquisition.  \textbf{Pengbo Wang:} Conceptualization,
Methodology, Formal analysis, Software, Validation, Writing--review \& editing.

\subsection*{Funding}
This work was supported by the National Natural Science Foundation of China
[grant number 72373170].

\subsection*{Declaration of competing interest}
The authors declare that they have no known competing financial interests or
personal relationships that could have appeared to influence the work reported
in this paper.

\subsection*{Declaration of generative AI and AI-assisted technologies in the
manuscript preparation process}
During the preparation of this work, the authors used OpenAI Codex to support
literature organization, language editing, and consistency checks. The authors
reviewed and edited the content as needed and take full responsibility for the
content of the publication.

\renewcommand{\bibfont}{\normalsize}
\setlength{\bibsep}{0pt}
\bibliographystyle{plainnat}
\addcontentsline{toc}{section}{References}
\bibliography{references}

@article{baik1997collective,
  title = {Collective Rent Seeking with Endogenous Group Sizes},
  author = {Baik, Kyung Hwan and Lee, Sanghack},
  journal = {European Journal of Political Economy},
  year = {1997},
  volume = {13},
  number = {1},
  pages = {121--130},
  doi = {10.1016/S0176-2680(96)00041-9}
}

@article{baik2001strategic,
  title = {Strategic Groups and Rent Dissipation},
  author = {Baik, Kyung Hwan and Lee, Sanghack},
  journal = {Economic Inquiry},
  year = {2001},
  volume = {39},
  number = {4},
  pages = {672--684},
  doi = {10.1093/ei/39.4.672}
}

@article{baik1995competitive,
  title = {Competitive-Share Group Formation in Rent-Seeking Contests},
  author = {Baik, Kyung Hwan and Shogren, Jason F.},
  journal = {Public Choice},
  year = {1995},
  volume = {83},
  number = {1--2},
  pages = {113--126},
  doi = {10.1007/BF01047687}
}

@article{chade2020competing,
  title = {Competing Teams},
  author = {Chade, Hector and Eeckhout, Jan},
  journal = {Review of Economic Studies},
  year = {2020},
  volume = {87},
  number = {3},
  pages = {1134--1173},
  doi = {10.1093/restud/rdz022}
}

@article{chowdhury2017small,
  title = {{``Small, yet Beautiful''}: Reconsidering the Optimal Design of Multi-Winner Contests},
  author = {Chowdhury, Subhasish M. and Kim, Sang-Hyun},
  journal = {Games and Economic Behavior},
  year = {2017},
  volume = {104},
  pages = {486--493},
  doi = {10.1016/j.geb.2017.05.012}
}

@article{eliaz2018simple,
  title = {A Simple Model of Competition Between Teams},
  author = {Eliaz, Kfir and Wu, Qinggong},
  journal = {Journal of Economic Theory},
  year = {2018},
  volume = {176},
  pages = {372--392},
  doi = {10.1016/j.jet.2018.04.006}
}

@article{hart1983endogenous,
  title = {Endogenous Formation of Coalitions},
  author = {Hart, Sergiu and Kurz, Mordecai},
  journal = {Econometrica},
  year = {1983},
  volume = {51},
  number = {4},
  pages = {1047--1064},
  doi = {10.2307/1912051}
}

@article{konishi2020sequential,
  title = {Sequential Formation of Alliances in Survival Contests},
  author = {Konishi, Hideo and Pan, Chen-Yu},
  journal = {International Journal of Economic Theory},
  year = {2020},
  volume = {16},
  number = {1},
  pages = {95--105},
  doi = {10.1111/ijet.12246}
}

@article{konishi2021endogenous,
  title = {Endogenous Alliances in Survival Contests},
  author = {Konishi, Hideo and Pan, Chen-Yu},
  journal = {Journal of Economic Behavior \& Organization},
  year = {2021},
  volume = {189},
  pages = {337--358},
  doi = {10.1016/j.jebo.2021.06.039}
}

@article{konishi2026formation,
  title = {Formation of Teams in Contests: Tradeoffs Between Inter- and Intra-Team Inequalities},
  author = {Konishi, Hideo and Pan, Chen-Yu and Simeonov, Dimitar},
  journal = {Social Choice and Welfare},
  year = {2026},
  doi = {10.1007/s00355-026-01663-9}
}

@article{moldovanu2006contest,
  title = {Contest Architecture},
  author = {Moldovanu, Benny and Sela, Aner},
  journal = {Journal of Economic Theory},
  year = {2006},
  volume = {126},
  number = {1},
  pages = {70--96},
  doi = {10.1016/j.jet.2004.10.004}
}

@article{morelli2016internal,
  title = {Internal Hierarchy and Stable Coalition Structures},
  author = {Morelli, Massimo and Park, In-Uck},
  journal = {Games and Economic Behavior},
  year = {2016},
  volume = {96},
  pages = {90--96},
  doi = {10.1016/j.geb.2016.01.007}
}

@article{sanchezpages2007endogenous,
  title = {Endogenous Coalition Formation in Contests},
  author = {S{\'a}nchez-Pag{\'e}s, Santiago},
  journal = {Review of Economic Design},
  year = {2007},
  volume = {11},
  number = {2},
  pages = {139--163},
  doi = {10.1007/s10058-007-0033-4}
}

@techreport{nevrekar2022efficiency,
  title = {Efficiency Effects on Coalition Formation in Contests},
  author = {Nevrekar, Saish},
  institution = {University of Zurich, Department of Economics},
  type = {Working Paper No.},
  year = {2022},
  number = {412},
  doi = {10.5167/uzh-218593},
  url = {https://doi.org/10.5167/uzh-218593}
}

@article{garfinkel2004stability,
  title = {On the Stability of Group Formation: Managing the Conflict Within},
  author = {Garfinkel, Michelle R.},
  journal = {Conflict Management and Peace Science},
  year = {2004},
  volume = {21},
  number = {1},
  pages = {43--68},
  doi = {10.1080/07388940490433918}
}

@article{bloch2006peace,
  title = {When Does Universal Peace Prevail? Secession and Group Formation in Conflict},
  author = {Bloch, Francis and S{\'a}nchez-Pag{\'e}s, Santiago and Soubeyran, Rapha{\"e}l},
  journal = {Economics of Governance},
  year = {2006},
  volume = {7},
  number = {1},
  pages = {3--29},
  doi = {10.1007/s10101-005-0099-9}
}

@article{cornes2005asymmetric,
  title = {Asymmetric Contests with General Technologies},
  author = {Cornes, Richard and Hartley, Roger},
  journal = {Economic Theory},
  year = {2005},
  volume = {26},
  number = {4},
  pages = {923--946},
  doi = {10.1007/s00199-004-0566-5}
}

@article{deng2025monopoly,
  title = {From Monopoly to Competition: When Do Optimal Contests Prevail?},
  author = {Deng, Xiaotie and Gafni, Yotam and Lavi, Ron and Lin, Tao and Ling, Hongyi},
  journal = {Games and Economic Behavior},
  year = {2025},
  volume = {153},
  pages = {268--293},
  doi = {10.1016/j.geb.2025.07.001}
}

@article{yamazaki2008asymmetric,
  title = {On the Existence and Uniqueness of Pure-Strategy {Nash} Equilibrium in Asymmetric Rent-Seeking Contests},
  author = {Yamazaki, Takeshi},
  journal = {Journal of Public Economic Theory},
  year = {2008},
  volume = {10},
  number = {2},
  pages = {317--327},
  doi = {10.1111/j.1467-9779.2008.00364.x}
}

@article{szidarovszky1997existence,
  title = {On the Existence and Uniqueness of Pure {Nash} Equilibrium in Rent-Seeking Games},
  author = {Szidarovszky, Ferenc and Okuguchi, Koji},
  journal = {Games and Economic Behavior},
  year = {1997},
  volume = {18},
  number = {1},
  pages = {135--140},
  doi = {10.1006/game.1997.0517}
}

@article{nitzan1991sharing,
  title = {Rent-Seeking with Non-Identical Sharing Rules},
  author = {Nitzan, Shmuel},
  journal = {Public Choice},
  year = {1991},
  volume = {71},
  number = {1--2},
  pages = {43--50},
  doi = {10.1007/BF00138449}
}

@incollection{ray2015coalition,
  title = {Coalition Formation},
  author = {Ray, Debraj and Vohra, Rajiv},
  booktitle = {Handbook of Game Theory with Economic Applications},
  editor = {Young, H. Peyton and Zamir, Shmuel},
  publisher = {Elsevier},
  year = {2015},
  volume = {4},
  pages = {239--326},
  doi = {10.1016/B978-0-444-53766-9.00005-7}
}

@article{ray1999theory,
  title = {A Theory of Endogenous Coalition Structures},
  author = {Ray, Debraj and Vohra, Rajiv},
  journal = {Games and Economic Behavior},
  year = {1999},
  volume = {26},
  number = {2},
  pages = {286--336},
  doi = {10.1006/game.1998.0648}
}

@article{yi1997stable,
  title = {Stable Coalition Structures with Externalities},
  author = {Yi, Sang-Seung},
  journal = {Games and Economic Behavior},
  year = {1997},
  volume = {20},
  number = {2},
  pages = {201--237},
  doi = {10.1006/game.1997.0567}
}

@article{bogomolnaia2002stability,
  title = {The Stability of Hedonic Coalition Structures},
  author = {Bogomolnaia, Anna and Jackson, Matthew O.},
  journal = {Games and Economic Behavior},
  year = {2002},
  volume = {38},
  number = {2},
  pages = {201--230},
  doi = {10.1006/game.2001.0877}
}

@article{bloch1996sequential,
  title = {Sequential Formation of Coalitions in Games with Externalities and Fixed Payoff Division},
  author = {Bloch, Francis},
  journal = {Games and Economic Behavior},
  year = {1996},
  volume = {14},
  number = {1},
  pages = {90--123},
  doi = {10.1006/game.1996.0043}
}

@article{dreze1980hedonic,
  title = {Hedonic Coalitions: Optimality and Stability},
  author = {Dr{\`e}ze, Jacques H. and Greenberg, Joseph},
  journal = {Econometrica},
  year = {1980},
  volume = {48},
  number = {4},
  pages = {987--1003},
  doi = {10.2307/1912943}
}

@article{andelman2009strong,
  title = {Strong Price of Anarchy},
  author = {Andelman, Nir and Feldman, Michal and Mansour, Yishay},
  journal = {Games and Economic Behavior},
  year = {2009},
  volume = {65},
  number = {2},
  pages = {289--317},
  doi = {10.1016/j.geb.2008.03.005}
}

@article{hafalir2007efficiency,
  title = {Efficiency in Coalition Games with Externalities},
  author = {Hafalir, Isa E.},
  journal = {Games and Economic Behavior},
  year = {2007},
  volume = {61},
  number = {2},
  pages = {242--258},
  doi = {10.1016/j.geb.2007.01.009}
}

@article{dutta1997stable,
  title = {Stable Networks},
  author = {Dutta, Bhaskar and Mutuswami, Suresh},
  journal = {Journal of Economic Theory},
  year = {1997},
  volume = {76},
  number = {2},
  pages = {322--344},
  doi = {10.1006/jeth.1997.2306}
}

@article{jackson1996strategic,
  title = {A Strategic Model of Social and Economic Networks},
  author = {Jackson, Matthew O. and Wolinsky, Asher},
  journal = {Journal of Economic Theory},
  year = {1996},
  volume = {71},
  number = {1},
  pages = {44--74},
  doi = {10.1006/jeth.1996.0108}
}

@article{lee2022group,
  title = {Group Formation in a Dominance-Seeking Contest},
  author = {Lee, Dongryul and Kim, Pilwon},
  journal = {Social Choice and Welfare},
  year = {2022},
  volume = {58},
  number = {1},
  pages = {39--68},
  doi = {10.1007/s00355-021-01346-7}
}

@article{dollinger2026alliances,
  title = {Alliances and Technological Partnerships in Contests},
  author = {Dollinger, J{\'e}r{\^o}me},
  journal = {Journal of Public Economic Theory},
  year = {2026},
  volume = {28},
  number = {2},
  pages = {e70107},
  doi = {10.1111/jpet.70107}
}

@article{sun2026alliance,
  title = {Alliance Formation in Team Contests with {Blotto} Budgets},
  author = {Sun, Xutang and Wang, Youyou and Zheng, Jie},
  journal = {Journal of Economic Behavior \& Organization},
  year = {2026},
  volume = {246},
  pages = {107541},
  doi = {10.1016/j.jebo.2026.107541}
}

@techreport{sakamoto2025selfenforcing,
  title = {Self-Enforcing Stable Sets},
  author = {Sakamoto, Hiroaki and Traeger, Christian P.},
  institution = {CESifo},
  type = {CESifo Working Paper},
  number = {12360},
  year = {2025},
  note = {Version of December 2025},
  url = {https://www.ifo.de/DocDL/cesifo1_wp12360.pdf}
}

\clearpage
\appendix
\part*{Appendices}
\addcontentsline{toc}{part}{Appendices}
\numberwithin{figure}{section}
\numberwithin{table}{section}
\numberwithin{theorem}{section}
\numberwithin{proposition}{section}
\numberwithin{corollary}{section}
\numberwithin{lemma}{section}
\numberwithin{definition}{section}
\numberwithin{remark}{section}

\section{Continuation contests and payoff accounting}\label{ec:continuation}

\subsection{The role lottery}

A team of size \(k\) receives total prize \(P_k\) if it wins.  Before effort is
chosen, one member is selected uniformly to be the active role.  The active
role has winning-state value \(a_k\leq P_k\), chooses team effort \(x\), and
pays \(c(x)\).  The remaining prize can be distributed among inactive members.
Ex ante symmetry of the role lottery implies that each member's payoff is
\begin{equation}\label{ec:eq:payoff}
 u_k=\frac{P_kp_k-c(x_k)}{k}.
\end{equation}
For the governance rule used in the paper,
\begin{equation}\label{ec:eq:role-value}
 a_k=P_k r_k,\qquad
 r_k=q+\frac{1-q}{k},\qquad q\in[0,1].
\end{equation}

\subsection{General interior continuation: attribution and coordinate map}

For the general mechanism, let team \(j\) have effective value \(a_j>0\)
and cost \(c_j\).  The smooth convex-cost class and zero-origin condition
stated in Section~\ref{subsec:continuation} are the present-coordinate version
of the interior asymmetric-contest case in
\citet[Proposition~1]{yamazaki2008asymmetric} and
\citet[Proposition~3(5)]{cornes2005asymmetric}.  In the production notation
used there, write output as \(x=f_j(e)\) and inverse technology as
\(c_j=f_j^{-1}\), after normalizing the valuation into either the production
function or the cost.  The condition \(c'_{j,+}(0)=0\) is the inverse form of
unbounded marginal product at the origin.  Heterogeneous \(a_j\) creates no
additional theorem: it can equivalently be absorbed by
\(\widehat c_j=c_j/a_j\).  Thus Lemma~\ref{lem:continuation} is an adaptation,
not a new existence or uniqueness result.

The scalar share argument verifies the mapping directly.  Given
\(X>0\), define \(p_j(X)\in(0,1)\) by
\begin{equation}\label{ec:eq:general-share}
 a_j(1-p_j)=Xc_j'(p_jX).
\end{equation}
At \(p=0\), the left side is \(a_j>0\) and the right side is zero; at
\(p=1\), the left side is zero and the right side is strictly positive.  The
difference is strictly decreasing because

\[
 -a_j-X^2c_j''(pX)<0.
\]

Hence \(p_j(X)\) exists and is unique.  Moreover,
\begin{equation}\label{ec:eq:general-share-derivative}
 p_j'(X)
 =-\frac{c_j'(p_jX)+p_jXc_j''(p_jX)}
          {a_j+X^2c_j''(p_jX)}<0.
\end{equation}
As \(X\downarrow0\), \(p_j(X)\to1\), so
\(\sum_jp_j(X)\to J>1\).  As \(X\to\infty\), \(p_j(X)\to0\): otherwise a
subsequence bounded away from zero would make the right side of
\eqref{ec:eq:general-share} diverge while the left side remains at most
\(a_j\).  Continuity and strict monotonicity therefore give a unique \(X>0\)
such that \(\sum_jp_j(X)=1\).

The corresponding efforts \(x_j=p_jX\) satisfy every first-order condition.
Given positive rival effort, each objective is strictly concave; an inactive
team has a positive right derivative at zero because \(c'_{j,+}(0)=0\).
The all-zero profile is ruled out under either convention stated in the main
text using only
\(c_j(0)=0\) and continuity.  These observations recover the unique interior
equilibrium asserted in Lemma~\ref{lem:continuation}.

\subsection{Unit effort cost}

Let \(X=\sum_jx_j\) and \(p_j=x_j/X\).  With \(c(x)=x\), the first-order
condition for an active team is
\[
 a_j(1-p_j)=X.
\]
If the active set \(\mathcal A\) contains \(m\geq2\) teams, summing
\(p_j=1-X/a_j\) gives
\begin{equation}\label{ec:eq:linear-active}
 X=\frac{m-1}{\sum_{j\in\mathcal A}1/a_j},
 \qquad
 p_j=1-\frac{X}{a_j},
 \qquad
 x_j=Xp_j.
\end{equation}
This candidate is an equilibrium precisely when \(a_j>X\) for active teams
and \(a_j\leq X\) for inactive teams.  Equality is assigned to the inactive
side.  The convention removes spurious zero-effort ``active'' teams without
changing effort or payoffs.

In the linear-benefit benchmark \(P_k=k\), so \(a_k=kr_k\).  Substituting
\(x_k=Xp_k=a_kp_k(1-p_k)\) into \eqref{ec:eq:payoff} yields
\begin{equation}\label{ec:eq:linear-member}
 U_k(p)=p\{1-r_k(1-p)\}.
\end{equation}
This expression is strictly increasing in \(p\) and weakly decreasing in
\(r_k\).

\subsection{Isoelastic and quadratic effort costs}

For \(c(x)=x^\rho/\rho\), \(\rho>1\), an interior team satisfies
\begin{equation}\label{ec:eq:isoelastic-foc}
 a_j(1-p_j)=X^\rho p_j^{\rho-1}.
\end{equation}
Since \(c'(0)=0\), every positive-value team is active.  For fixed
\(\lambda=X^\rho\), the equation has one solution \(p_j\in(0,1)\), and
\(\sum_jp_j=1\) determines a unique \(\lambda\).

When \(\rho=2\), write \(\lambda=X^2\).  Then
\begin{equation}\label{ec:eq:quadratic-active}
 p_j=\frac{a_j}{a_j+\lambda},
 \qquad
 \sum_j\frac{a_j}{a_j+\lambda}=1.
\end{equation}
The left side is continuous and strictly decreasing from the number of teams
to zero, so the positive solution is unique.

\section{Proofs for clone deviations}\label{ec:splitclone}

\begin{lemma}[Probability monotonicity]\label{ec:lem:p-monotone}
Under the conditions of Lemma~\ref{lem:continuation}, fix a focal team of type
\(\tau=(k_\tau,P_\tau,a_\tau,c_\tau)\).  Holding its primitives fixed,
its equilibrium winning probability is strictly decreasing in aggregate
effort \(X\).
\end{lemma}

\begin{proof}
Write
\[
 F(p,X)=a_\tau(1-p)-Xc_\tau'(pX)=0.
\]
Then
\[
 F_p=-a_\tau-X^2c_\tau''(pX)<0,\qquad
 F_X=-c_\tau'(pX)-pXc_\tau''(pX)<0.
\]
The implicit-function theorem gives
\[
 \frac{dp}{dX}=-\frac{F_X}{F_p}<0.
\]
\end{proof}

\begin{lemma}[Member-payoff monotonicity]\label{ec:lem:u-monotone}
Under the conditions of Lemma~\ref{ec:lem:p-monotone}, suppose also that
\(a_\tau\leq P_\tau\).  Each focal-team member's ex ante payoff is strictly
decreasing in \(X\).
\end{lemma}

\begin{proof}
From \(k_\tau u_\tau=P_\tau p-c_\tau(pX)\),
\[
 k_\tau\frac{du_\tau}{dX}
 =[P_\tau-Xc_\tau'(pX)]\frac{dp}{dX}-pc_\tau'(pX).
\]
The first-order condition gives
\[
 Xc_\tau'(pX)=a_\tau(1-p)\leq a_\tau\leq P_\tau.
\]
The first bracket is nonnegative, \(dp/dX<0\), and the last term is strictly
negative.  Hence \(du/dX<0\).
\end{proof}

\begin{proof}[Proof of Theorem~\ref{thm:splitclone}]
At a given aggregate effort, all \(\tau\)-teams solve the same strictly
monotone scalar equation, so their members share the same original payoff
\(u_\tau(X(\pi))\).  Lemma~\ref{lem:continuation} makes that payoff
single-valued for each architecture.  In the split, every coalition member
receives the clone payoff \(u_\tau(X(\pi'))\).  Lemma
\ref{ec:lem:u-monotone} makes the two rankings strict and opposite.
\end{proof}

\begin{proof}[Proof of Corollary~\ref{cor:plateau}]
For each \(\pi'\in\mathcal C_\tau(\pi)\cap\Pi\), optimality gives
\(X(\pi')\leq X(\pi)\).  Stability and
Theorem~\ref{thm:splitclone} give the reverse weak inequality.  Hence
equality holds.  If this intersection is nonempty, choose one such \(\pi'\).
The clone has size \(k_\tau\) and draws positive numbers of members from at
least two size-\(k_\tau\) sources.  No source can be exhausted, since that
would use all \(k_\tau\) clone positions and leave no positive extraction from
a second source.  Thus every source leaves a nonempty residual while the clone
adds one team.  The number of teams, and therefore the team-size multiset,
changes; strict optimality gives \(X(\pi')<X(\pi)\), contradicting plateau
equality and yielding a block.  The stated \(1+(k-1)\) split is the canonical
homogeneous instance.
\end{proof}

The common-source-payoff condition is tight.  The reverse implication fails
with heterogeneous source payoffs.  Consider
\(N=4\), capacity \(n=2\), and quadratic cost \(c(x)=x^2/2\).  Before the
deviation, two size-two teams have \((a,P)=(1,1)\) and \((2,2)\).  Draw one
member from each, form a size-two clone of the first type, and assign each
singleton residual \((a,P)=(1/10,1/10)\).  This example uses
identity-dependent primitives rather than the size-dependent benchmark.
Direct solution gives
\[
 X(\pi)=2^{1/4}>X(\pi')=\frac1{\sqrt2},
 \qquad u_{\mathrm{clone}}(\pi')=\frac5{18}.
\]
Relative to the type-A source member, the gain is
\(-2/9+\sqrt2/4>0\), whereas relative to the type-B source member it is
\(-67/18+5\sqrt2/2<0\).  Lower effort therefore does not make every
heterogeneous source member better off.  Thus aggregate effort alone does not
order every heterogeneous source member's gain.

\section{Proofs for the exact linear benchmark}\label{app:linear-proofs}

\subsection{Stability boundary}\label{ec:linear-stability}

Throughout this appendix, let
\begin{equation}\label{ec:eq:linear-benchmark}
 P_k=k,\qquad a_k=1+q(k-1),\qquad c(x)=x,
 \qquad A=a_n.
\end{equation}
The target effort and each target team's winning probability are
\begin{equation}\label{ec:eq:linear-target}
 X_0=\frac{M-1}{M}A,\qquad p_0=\frac1M.
\end{equation}

\subsubsection{All deviations below the candidate boundary}

A nontrivial one-new-team coalition extracts
\[
 t_1,\ldots,t_\ell\geq1,\qquad
 s=\sum_{i=1}^{\ell}t_i\leq n,
\]
and leaves residual teams of sizes \(n-t_i\).  A whole-team transfer
\((\ell,s,t_1)=(1,n,n)\) only relabels an incumbent and is excluded.

Define the linear residual-loss function
\begin{equation}\label{ec:eq:linear-g}
 g_q^{\mathrm L}(t)=\frac{qt}{A-qt}.
\end{equation}
The candidate stability threshold is the positive root of
\begin{equation}\label{ec:eq:linear-qB-poly}
 F_{M,n}(q)
 =(M-1)(n-1)(n-2)q^2+[n(M-2)+2]q-1=0,
\end{equation}
namely
\begin{equation}\label{ec:eq:linear-qB}
 \qB=
 \frac{2}{
 \sqrt{[n(M-2)+2]^2+4(M-1)(n-1)(n-2)}
 +n(M-2)+2}.
\end{equation}
Direct substitution into \(F_{M,n}\) shows
\begin{equation}\label{ec:eq:active-bound}
 \qB<\frac{1}{(M-1)(n-1)},\qquad
 q\leq\qB\ \Longrightarrow\ A<\frac{M}{M-1}.
\end{equation}

\begin{lemma}[All-active region]\label{ec:lem:linear-all-active}
Every team in every nontrivial one-new-team continuation is active when
\(q\leq\qB\).
\end{lemma}

\begin{proof}
A nontrivial deviation leaves \(M+1\) nonempty teams, with values in
\([1,A]\).  In the all-active candidate, fix a team with value \(d\).  Using
\eqref{ec:eq:linear-active} and bounding the other \(M\) values by \(A\),
\[
 X\leq\frac{M}{1/d+M/A}=\frac{MAd}{A+Md}<d,
\]
where the final inequality follows from \((M-1)A<Md\) and \(d\geq1\).
Thus the candidate satisfies every activity inequality.
\end{proof}

For an all-active deviation, multiply the reciprocal-value sum in
\eqref{ec:eq:linear-active} by the new team's value \(a_s\).  Its winning
probability can be written
\begin{equation}\label{ec:eq:linear-deviation-p}
 p_s=1-\frac{M}{D},\qquad
 D=1+\frac{a_s}{A}\left[M+\sum_{i=1}^{\ell}g_q^{\mathrm L}(t_i)\right].
\end{equation}

\begin{proposition}[One-source protection]\label{ec:prop:one-source}
A nontrivial coalition recruiting from only one full-cap incumbent cannot
strictly block.
\end{proposition}

\begin{proof}
The deviation creates a size-\(s<n\) team and a positive residual from one
incumbent while \(M-1\) full teams remain.  Each untouched full team has value
at least that of the new team and therefore has equilibrium probability at
least \(p_s\).  If the residual is active, its strictly positive probability
and adding up imply \(1>Mp_s\).  If it is inactive, then \(q>0\): at \(q=0\),
all \(M+1\) nonempty teams have the same value and are active.  Since \(s<n\),
\(q>0\) makes every untouched full team strictly more valuable than the new
team and hence gives it probability strictly above \(p_s\); adding up again
implies \(1>Mp_s\).  Thus \(p_s<1/M\) in either case.  Together with
\(r_s\geq r_n\), \eqref{ec:eq:linear-member} rules out a strict gain.
\end{proof}

\begin{lemma}[Canonical envelope]\label{ec:lem:linear-envelope}
For positive integers \(t_i\) with at least two parts and
\(\sum_it_i\leq n\),
\[
 \sum_i g_q^{\mathrm L}(t_i)\leq
 g_q^{\mathrm L}(1)+g_q^{\mathrm L}(n-1).
\]
Consequently, the canonical extraction \((1,n-1)\) weakly maximizes the
replacement team's probability throughout the all-active region.
\end{lemma}

\begin{proof}
The discrete increments
\(g_q^{\mathrm L}(t+1)-g_q^{\mathrm L}(t)\) are increasing in \(t\).
Holding a sum fixed, transferring one unit from the smaller of two positive
parts to the larger therefore weakly raises their total \(g_q^{\mathrm L}\).
Repetition reduces any composition to two parts and then to \((1,s-1)\).
Because \(g_q^{\mathrm L}\) is increasing, the resulting value is maximized
at \(s=n\).  Finally \(a_s/A\leq1\) in
\eqref{ec:eq:linear-deviation-p}, with equality for
the canonical full-size replacement.
\end{proof}

\subsubsection{Closing the deviation envelope}

For the canonical full-size replacement, the new team has the target's
\(r_n\).  It is indifferent exactly when its probability is \(1/M\).
Substituting \((1,n-1)\) into \eqref{ec:eq:linear-deviation-p} and clearing
positive denominators gives \(F_{M,n}(q)=0\).

\begin{proof}[Proof of Theorem~\ref{thm:linear-stability}]
Proposition~\ref{ec:prop:one-source} handles deviations from one source.  For
deviations drawing from at least two sources and \(q\leq\qB\),
Lemmas~\ref{ec:lem:linear-all-active} and
\ref{ec:lem:linear-envelope} imply \(p_s\leq p^{\rm can}\leq1/M\).
Moreover \(s\leq n\) gives \(r_s\geq r_n\).  Since
\eqref{ec:eq:linear-member} is increasing in \(p\) and decreasing in \(r\),
\[
 U_s(p_s)\leq U_s(1/M)\leq U_n(1/M),
\]
so no coalition strictly benefits all its members.  At \(q=\qB\), equality is
attained by the canonical coalition.

For \(q>\qB\), its full-size clone has probability above \(1/M\) and therefore
strictly higher member payoff.  Before any residual-team activity transition,
this follows directly from the sign of \(F_{M,n}\).  Within a fixed active-set
region, normalize every active value by the clone value \(A\).  For a residual
of size \(n-t\),
\[
 \frac{a_{n-t}}{A}
 =\frac{1+q(n-t-1)}{1+q(n-1)}
\]
is weakly decreasing in \(q\), while every full-size ratio remains one.
Hence \(\sum_{j\in\mathcal A}A/a_j\) weakly rises and
\[
 p^{\rm can}=1-\frac{|\mathcal A|-1}
 {\sum_{j\in\mathcal A}A/a_j}
\]
does not fall.  At an activity transition, effort and clone payoff are
continuous under the equality-inactive convention, and removing a residual
rival weakly raises the clone's probability.  The strict block therefore
persists for every larger \(q\).
\end{proof}

\subsection{Global boundary}\label{ec:linear-global}

\subsubsection{Fixed active-team count}

Let \(N=Mn\) and
\[
 f_q(k)=\frac{1}{1+q(k-1)}.
\]
For \(q>0\), \(f_q\) is strictly decreasing and strictly convex.  If an
architecture has \(m\) active teams with sizes \(k_1,\ldots,k_m\), then
\[
 X=\frac{m-1}{\sum_{i=1}^{m}f_q(k_i)}.
\]
When \(m\leq M\), capacity gives
\begin{equation}\label{ec:eq:m-small}
 X\leq\frac{m-1}{m}a_n\leq\frac{M-1}{M}a_n=X_0.
\end{equation}
When \(m\geq M\), assigning all \(N\) contestants to the active teams and
balancing their integer sizes minimizes the reciprocal-value sum and therefore
gives an upper bound.  Set
\begin{equation}\label{ec:eq:balanced-m}
 d_m=\left\lfloor\frac{N}{m}\right\rfloor,\qquad
 r_m=N-md_m,
\end{equation}
\[
 D_m=\frac{m-r_m}{a_{d_m}}+\frac{r_m}{a_{d_m+1}},
 \qquad B_m=\frac{m-1}{D_m}.
\]
Then every architecture with \(m\) active teams satisfies \(X\leq B_m\).
Equality requires the near-balanced integer partition and no inactive
contestants.

\subsubsection{First-competitor principle}

Linearly interpolate \(f_q\) between integer team sizes and call the resulting
convex function \(\bar f_q\).  The function
\begin{equation}\label{ec:eq:perspective}
 D(m)=m\bar f_q(N/m)
\end{equation}
is a perspective of \(\bar f_q\), hence is convex in \(m\).  On integer
arguments it equals \(D_m\).  Let
\[
 \Delta_m=D_{m+1}-D_m,\quad
 C_m=\frac{D_m}{m-1}=\frac1{B_m},\quad
 H_m=(m-1)\Delta_m-D_m.
\]
Convexity makes \(\Delta_m\) nondecreasing, and direct algebra gives
\begin{equation}\label{ec:eq:H}
 C_{m+1}-C_m=\frac{H_m}{m(m-1)},\qquad
 H_{m+1}-H_m=m(\Delta_{m+1}-\Delta_m)\geq0.
\end{equation}
Therefore
\begin{equation}\label{ec:eq:first-competitor}
 B_M\geq B_{M+1}
 \quad\Longrightarrow\quad
 B_M\geq B_m\quad\text{for every }m>M.
\end{equation}
This is the first-competitor principle: once the target beats the best
\((M+1)\)-team partition, it beats every partition with still more active
teams.  Equation \eqref{ec:eq:m-small} handles fewer active teams.

\subsubsection{Threshold and uniqueness}

For \(m=M+1\), write
\begin{equation}\label{ec:eq:d-r}
 d=\left\lfloor\frac{Mn}{M+1}\right\rfloor,\qquad
 r=Mn-(M+1)d.
\end{equation}
The near-balanced competitor contains \(M+1-r\) teams of size \(d\) and
\(r\) teams of size \(d+1\).  The tie \(B_{M+1}=X_0\) is
\begin{equation}\label{ec:eq:G}
 G_{M,n}(q)
 =M^2a_da_{d+1}
 -(M-1)a_n\big[(M+1-r)a_{d+1}+ra_d\big]=0.
\end{equation}
Writing \(G=1+\gamma_1q+\gamma_2q^2\),
\begin{align}
 \gamma_1={}&M^2(2d-1)
 -(M-1)\big[Mn-2r+(M+1)(n-1)\big],\\
 \gamma_2={}&M^2d(d-1)
 -(M-1)(n-1)(Mn-2r).
\end{align}
To verify the sign of the quadratic term, set \(t=n-d\), so that
\(r=Mt-d\).  Direct rearrangement gives
\[
 -\gamma_2=(M-2)d(d-1)
 +(M-1)t\big[M(t+1)-2r\big]>0.
\]
If \(t=1\), then \(r\leq M-1\); if \(t\geq2\), then \(r\leq M\) and
\(M(t+1)-2r\geq M(t-1)>0\).  Hence \(\gamma_2<0\).

Moreover, \(G(0)=1\).  At \(q=1\), Cauchy's inequality gives
\[
 B_{M+1}(1)\leq\frac{MN}{(M+1)^2}
 <\frac{N(M-1)}{M^2}=X_0(1),
\]
where the strict inequality is equivalent to \(M^2-M-1>0\).  Thus
\(G(1)<0\).  Since the product of the two roots is \(1/\gamma_2<0\), exactly
one root is positive, and the endpoint signs place it in \((0,1)\):
\begin{equation}\label{ec:eq:linear-qG}
 \qG=\frac{2}{\sqrt{\gamma_1^2-4\gamma_2}-\gamma_1}.
\end{equation}

The near-balanced competitor is all-active throughout the comparison up to
this root.  This is immediate when \(r=0\).  For \(r>0\), let
\(L=a_d\).  Its all-active effort satisfies
\[
 B_{M+1}(q)<L \quad\Longleftrightarrow\quad q(r-d)<1.
\]
If \(r\leq d\), the inequality holds on \([0,1]\).  If \(r>d\), set
\(q_a=1/(r-d)\).  At \(q_a\), \(B_{M+1}=L\), while
\[
 X_0(q_a)-L
 =\frac{(M-1)(n-1-d)+M-r}{M(r-d)}>0.
\]
Consequently \(G(q_a)<0\), so the unique positive crossing satisfies
\(\qG<q_a\).  Hence every team in the competitor is active for all
\(q\leq\qG\).

\begin{proof}[Proof of Theorem~\ref{thm:linear-global}]
For \(q<\qG\), the displayed competitor has effort above \(X_0\).  For
\(q\geq\qG\), it has effort weakly below \(X_0\); the first-competitor
principle and \eqref{ec:eq:m-small} bound every possible active set.
Strict convexity identifies the unique fixed-\(m\) maximizer, and strictness in
\eqref{ec:eq:H} after the threshold rules out all \(m\neq M\).  At \(m=M\),
capacity and \(N=Mn\) force the target.  The \(q=0\) endpoint follows directly
from \eqref{ec:eq:linear-active}.
\end{proof}

\begin{proof}[Proof of Theorem~\ref{thm:linear-gap}]
At \(q=\qB\), the canonical full-size clone has probability \(1/M\), hence its
split architecture has the same effort \(A(1-1/M)=X_0\) as the target.  That
architecture has \(M+1\) teams.  Among partitions with that team count, strict
convexity of \(f_q\) makes the near-balanced partition weakly better, and
strictly better unless the canonical size vector is itself near-balanced.
The latter occurs exactly when \(n=2\).  Hence the target cannot reach the
global boundary before \(\qB\), and the inequality is strict for \(n>2\).
\end{proof}

\begin{figure}
\centering
\includegraphics[width=.92\textwidth]{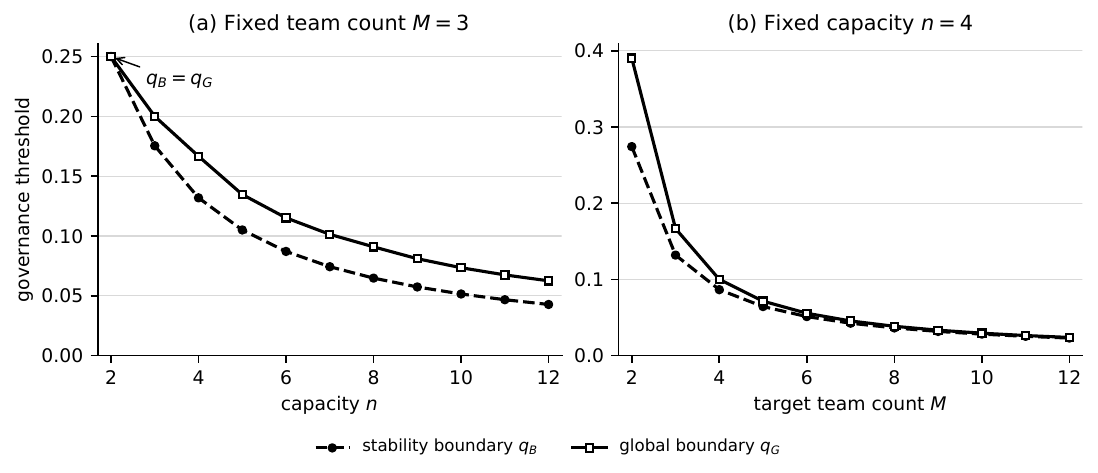}
\caption{Exact linear thresholds.  Dashed circles show the largest incentive
level \(q_B\) at which the full-cap target is one-new-team stable; solid
squares show the smallest level \(q_G\) at which it is a weak global
aggregate-effort maximizer.  Panel (a) varies capacity at \(M=3\); the
boundaries coincide at \(n=2\) and separate for every displayed \(n>2\).
Panel (b) varies the number of target teams at \(n=4\), where the strict gap
persists throughout.  Values use the exact expressions in
\eqref{eq:qB-def} and \eqref{ec:eq:linear-qG}.}
\label{fig:linear-exact-thresholds}
\end{figure}
\FloatBarrier

\section{Quadratic deviation reduction and stability}\label{ec:quadratic-stability}

Set
\begin{equation}\label{ec:eq:quadratic-benchmark}
 c(x)=\frac{x^2}{2},\qquad
 P_k=k^2,\qquad
 a_k=k[1+q(k-1)].
\end{equation}
This is the matched isoelastic case \(\rho=\beta=2\).

\subsection{Why full-size cross-source replacements suffice}

For general isoelastic cost, implicit differentiation of
\eqref{ec:eq:isoelastic-foc} gives
\[
 \varepsilon_i=
 \frac{p_i(1-p_i)}{(\rho-1)-(\rho-2)p_i}>0,
\]
\[
 \frac{\partial p_i}{\partial\ln a_i}
 =\varepsilon_i\left(1-\frac{\varepsilon_i}{\sum_h\varepsilon_h}\right)>0,
 \qquad
 \frac{\partial p_i}{\partial\ln a_j}
 =-\frac{\varepsilon_i\varepsilon_j}{\sum_h\varepsilon_h}<0
 \quad(j\neq i).
\]
Thus a team's probability rises with its own value and when a rival's value
falls.  Its member payoff is
\begin{equation}\label{ec:eq:isoelastic-payoff}
 U_k(p)=\frac{P_kp-a_kp(1-p)/\rho}{k}.
\end{equation}
For the role-lottery family with \(P_k=Vk^\beta\), \(V>0\), the effective
value is \(a_k=Vk^{\beta-1}(1-q+qk)\), which is nondecreasing in \(k\) when
\(\beta\geq1\).  Write \(h=(1-p)/\rho\in[0,1)\).  Then
\[
 U_k(p)=Vp\{(1-qh)k^{\beta-1}-(1-q)h k^{\beta-2}\}.
\]
After continuous interpolation in \(k\),
\[
 \frac{\partial U_k(p)}{\partial k}
 =Vp k^{\beta-3}
 \{(\beta-1)(1-qh)k-(\beta-2)(1-q)h\}.
\]
The expression in braces is nonnegative for \(1\leq\beta\leq2\).  For
\(\beta\geq2\), it is bounded below, using \(k\geq1\), by
\[
 (\beta-1)-(\beta-2+q)h\geq(\beta-1)(1-h)>0.
\]
Thus \(U_k(p)\) is nondecreasing in \(k\) for every fixed \(p\) whenever
\(\beta\geq1\).  It is also strictly increasing in \(p\), because, with
\(r_k=a_k/P_k\leq1\),
\[
 \frac{\partial U_k(p)}{\partial p}
 =Vk^{\beta-1}\left(1-\frac{r_k}{\rho}
 +\frac{2r_kp}{\rho}\right)>0.
\]

\begin{lemma}[Full-size completion]\label{ec:lem:completion}
Under common isoelastic cost and the role-lottery family
\(P_k=Vk^\beta\), \(V>0\), \(\beta\geq1\), any profitable cross-source
one-new-team deviation of size \(c<n\) is weakly dominated by a profitable
full-size replacement using the same sources.
\end{lemma}

\begin{proof}
Add to the coalition one nonmember from a source it already uses.  The new
team's size and effective value weakly rise, while one rival residual shrinks
or disappears.  Its probability weakly rises by the comparative statics above,
and its payoff at a fixed probability does not fall with size.  Repeat until
the new team reaches \(n\).  The original strict payoff gain is preserved, and
the added homogeneous members receive the same final payoff.
\end{proof}

\begin{lemma}[One-source protection]\label{ec:lem:quad-one-source}
Under the same conditions, no nontrivial one-source split strictly blocks a
full-cap balanced target.
\end{lemma}

\begin{proof}
After a one-source split there are \(M-1\) untouched full teams, the new
size-\(c<n\) team, and a positive residual.  The untouched teams have weakly
higher values and hence weakly higher probabilities than the new team.  The
residual has strictly positive probability, so the new team has probability
below \(1/M\).  Payoff increases in probability and is nondecreasing in team
size; therefore its members receive less than a target member.
\end{proof}

\subsection{Residual-loss superadditivity}

Let
\begin{equation}\label{ec:eq:quad-HC}
 H_q(t)=t[1+q(2n-t-1)],\qquad
 C_q=Mn[1+q(n-1)],
\end{equation}
\begin{equation}\label{ec:eq:quad-g}
 g_q(t)=\frac{(M-1)H_q(t)}{C_q-H_q(t)}.
\end{equation}
For a full-size replacement with extraction vector
\(t_1+\cdots+t_\ell=n\), the split--clone theorem and the target-intensity
calculation in Section~\ref{ec:quadratic-global} imply that it blocks exactly
when
\[
 \sum_{i=1}^{\ell}g_q(t_i)>1.
\]

For the canonical pair, clearing denominators in
\(g_q(1)+g_q(n-1)=1\) gives
\[
 A_2q^2+B_2q-(2M-1)=0,
\]
where
\begin{equation}\label{ec:eq:A2B2}
 A_2=2n(M-1)^2(n-1),\qquad
 B_2=2M^2n-6Mn+4M+3n-2.
\end{equation}
Its positive root is
\begin{equation}\label{ec:eq:qB2}
 \qBt=
 \frac{2(2M-1)}
 {B_2+\sqrt{B_2^2+4A_2(2M-1)}}.
\end{equation}

\begin{lemma}[Quadratic residual-loss superadditivity]\label{ec:lem:superadd}
For \(q\leq\qBt\) and positive integers \(a,b\) with \(a+b\leq n\),
\[
 g_q(a+b)\geq g_q(a)+g_q(b).
\]
At \(q=\qBt>0\), equality holds only for
\(\{a,b\}=\{1,n-1\}\).
\end{lemma}

\begin{proof}
Put \(s=a+b\), \(p=ab\), and \(A_0=1+q(2n-1)\), so that
\(H_q(t)=t(A_0-qt)\).  Direct common-denominator expansion gives
\begin{equation}\label{ec:eq:Xi-factor}
 g_q(a+b)-g_q(a)-g_q(b)
 =
 \frac{(M-1)ab\,\Xi_q(s,p)}
 {[C_q-H_q(a)][C_q-H_q(b)][C_q-H_q(s)]},
\end{equation}
where all three denominator factors are positive and
\begin{equation}\label{ec:eq:Xi}
 \Xi_q(s,p)=
 (A_0^2-qA_0s+q^2p)(2C_q-A_0s+qs^2)-2qC_q^2.
\end{equation}
For fixed \(s\),
\[
 \frac{\partial\Xi_q(s,p)}{\partial p}
 =q^2[2C_q-A_0s+qs^2]
 =q^2[2MA-H_q(s)]\geq0.
\]
For positive integer parts, \(p=ab\geq s-1\), with equality only at
\(\{a,b\}=\{1,s-1\}\).  Hence the extreme split minimizes the margin.

Let \(\underline\Xi_q(s)=\Xi_q(s,s-1)\) and \(w=n-s\).  Differentiating its
continuous extension gives
\[
 \frac{d\underline\Xi_q(s)}{ds}
 =-[1+2q(n-1)]Q_q(s),
\]
where
\begin{align*}
 Q_q(s)={}&1+(2Mn-1)q+n[2M(n-1)-n]q^2\\
 &+[4q+2(n-1)q^2]w+3q^2w^2.
\end{align*}
Every displayed coefficient is positive for \(M,n\geq2\), since
\(2M(n-1)-n>0\).  Thus \(\underline\Xi_q(s)\) is strictly decreasing on
\(2\leq s\leq n\), and the unique hardest endpoint is
\((a,b)=(1,n-1)\).  At that endpoint,
\(\Xi_q(n,n-1)\geq0\) is equivalent to
\(g_q(1)+g_q(n-1)\leq1\), which after expansion is
\[
 A_2q^2+B_2q-(2M-1)\leq0.
\]
This is equivalent to \(q\leq\qBt\).  Strictness in the product and
sum reductions gives the stated equality condition.
\end{proof}

\begin{proof}[Proof of Proposition~\ref{prop:quadratic-exact}, stability part]
Repeated application of Lemma~\ref{ec:lem:superadd} gives
\[
 \sum_i g_q(t_i)\leq g_q\!\left(\sum_it_i\right)=g_q(n)=1.
\]
Hence no full-size replacement blocks for \(q\leq\qBt\).  The equality
conditions identify the canonical pair.  Above the threshold that pair
strictly blocks.  Lemmas~\ref{ec:lem:completion} and
\ref{ec:lem:quad-one-source} extend the conclusion to the complete
one-new-team deviation class.
\end{proof}

\section{Quadratic global architecture choice}\label{ec:quadratic-global}

\subsection{Target intensity and deficit duality}

Let
\[
 A=a_n=n[1+q(n-1)],\qquad
 \lambda_0=(M-1)A.
\]
At the target intensity, a size-\(k\) team's implied probability is
\[
 p_k(\lambda_0)=\frac{a_k}{a_k+(M-1)A}.
\]
Because the sum of implied probabilities is strictly decreasing in
\(\lambda\), for any architecture \(\pi\),
\begin{equation}\label{ec:eq:intensity-test}
 X(\pi)>X(\target)
 \quad\Longleftrightarrow\quad
 \sum_{k\in\pi}p_k(\lambda_0)>1.
\end{equation}

Write a team's capacity deficit as \(t=n-k\).  Since
\[
 A-a_{n-t}=H_q(t),\qquad C_q=MA,
\]
direct calculation gives
\begin{equation}\label{ec:eq:p-g}
 p_{n-t}(\lambda_0)=\frac{1-g_q(t)}{M}.
\end{equation}
An architecture with \(M+L\) teams has deficits satisfying
\(\sum_it_i=nL\).  Equations \eqref{ec:eq:intensity-test} and
\eqref{ec:eq:p-g} yield the exact duality
\begin{equation}\label{ec:eq:deficit-duality}
 X(\pi)\leq X(\target)
 \quad\Longleftrightarrow\quad
 \sum_i g_q(t_i)\geq L.
\end{equation}

\subsection{The near-balanced candidate}

Let \(J=M+1\) and divide
\begin{equation}\label{ec:eq:division}
 n=Ju+v,\qquad 0\leq v\leq M.
\end{equation}
The near-balanced \(J\)-team partition has deficits \(u\) repeated \(J-v\)
times and \(u+1\) repeated \(v\) times.  Define \(\qGt\) as the unique
solution \(q\in(0,1)\) of
\begin{equation}\label{ec:eq:qG2-tie}
 (J-v)g_q(u)+v g_q(u+1)=1.
\end{equation}
To establish existence and uniqueness, observe that for \(0<t<n\),
\[
 \frac{\partial g_q(t)}{\partial q}
 =\frac{M(M-1)nt(n-t)}{[C_q-H_q(t)]^2}>0.
\]
At \(q=0\), \(g_0(t)=(M-1)t/(Mn-t)<t/n\) for every positive deficit
in the candidate, so their sum is below one.  At \(q=1\), put
\(x_i=(n-t_i)/n\).  Equation \eqref{ec:eq:p-g} gives
\[
 p_i=\frac{x_i^2}{x_i^2+M-1}\leq\frac{x_i}{M},
\]
with strict inequality for every nonfull team.  Since the candidate has
\(M+1\) teams, \(\sum_i x_i=M\), and at least one team is nonfull, its
probabilities sum to less than one; equivalently, its deficits satisfy
\(\sum_i g_1(t_i)>1\).  Continuity and the strict monotonicity above therefore
give exactly one solution in \((0,1)\).
Below this value the candidate beats the target.  The remaining task is to
show that at the tie no other finite partition beats it.

\subsection{A discrete supporting cubic}

Define
\begin{equation}\label{ec:eq:h}
 h_q(t)=\frac{t}{n}-g_q(t).
\end{equation}
Because \(\sum_it_i/n=L\), condition \eqref{ec:eq:deficit-duality} is
equivalent to
\begin{equation}\label{ec:eq:h-condition}
 \sum_i h_q(t_i)\leq0.
\end{equation}
At \(q=\qGt\), the two rounding points satisfy
\begin{equation}\label{ec:eq:h-round}
 (J-v)h_q(u)+v h_q(u+1)=0.
\end{equation}

Normalize
\[
 \gamma=\frac{nq}{1-q},\qquad y=\frac{t}{n},
\]
and set
\[
 a=\frac1\gamma-(M-2),\qquad
 D_\gamma(y)=(1-y)[1+\gamma(1-y)]+(M-1)(1+\gamma).
\]
At the relevant positive threshold, direct factorization gives
\begin{equation}\label{ec:eq:h-factor}
 h_\gamma(y)=
 \frac{\gamma y(1-y)(a-y)}{D_\gamma(y)},
 \qquad D_\gamma'(y)=-[1+2\gamma(1-y)]<0.
\end{equation}

Let \(y_0=u/n\), \(y_1=(u+1)/n\), \(w_i=y_i(1-y_i)\), and define the
arithmetic support root
\begin{equation}\label{ec:eq:abar}
 \bar a=
 \frac{(J-v)w_0y_0+vw_1y_1}
 {(J-v)w_0+vw_1}.
\end{equation}
Equivalently, with \(\theta=n\bar a\),
\begin{equation}\label{ec:eq:theta}
 \theta=
 \frac{(J-v)u^2(n-u)+v(u+1)^2(n-u-1)}
 {(J-v)u(n-u)+v(u+1)(n-u-1)}.
\end{equation}
In the nondegenerate rounding case, \eqref{ec:eq:h-round} expresses \(a\) as
a weighted average of \(y_0,y_1\) with the weights in
\eqref{ec:eq:abar} divided by \(D_\gamma(y_i)\).  Since
\(D_\gamma(y_0)>D_\gamma(y_1)\),
\begin{equation}\label{ec:eq:a-order}
 y_0<\bar a\leq a<y_1.
\end{equation}
Equality \(a=\bar a\) covers \(v=0\) and \(u=0\).

\begin{lemma}[Discrete supporting cubic]\label{ec:lem:supporting-cubic}
There is a constant \(\kappa>0\) such that, for every integer
\(t\in\{0,\ldots,n\}\),
\begin{equation}\label{ec:eq:supporting-cubic}
 h_q(t)\leq
 \kappa\,\frac{t}{n}\left(1-\frac{t}{n}\right)
 \left(\bar a-\frac{t}{n}\right)
\end{equation}
at \(q=\qGt\).  Equality holds at the two effective rounding points.
\end{lemma}

\begin{proof}
Let \(\phi_{\bar a}(y)=y(1-y)(\bar a-y)\).  Away from its roots,
\[
 R(y)=\frac{h_\gamma(y)}{\phi_{\bar a}(y)}
 =\frac{\gamma(a-y)}{D_\gamma(y)(\bar a-y)}.
\]
On \(y<\bar a\) and \(y>a\),
\[
 \frac{R'(y)}{R(y)}
 =\frac{a-\bar a}{(a-y)(\bar a-y)}
 -\frac{D_\gamma'(y)}{D_\gamma(y)}>0.
\]
Both \(a\) and \(\bar a\) lie in the grid gap \((y_0,y_1)\).
Equation \eqref{ec:eq:h-round} and the defining identity for \(\bar a\) imply
\(R(y_0)=R(y_1)=\kappa\).  Monotonicity then gives
\(R(t/n)\leq\kappa\) on the positive side of \(\phi\) and
\(R(t/n)\geq\kappa\) on its negative side, which yields the same upper bound
in both cases.  When \(a=\bar a\), use
\(R(y)=\gamma/D_\gamma(y)\); endpoint cases follow by continuity and direct
substitution.
\end{proof}

\subsection{Rounded cubic packing}

Define
\begin{equation}\label{ec:eq:psi}
 \psi_\theta(t)=t(n-t)(t-\theta).
\end{equation}

\begin{lemma}[Rounded cubic packing]\label{ec:lem:rounded-cubic}
For integers \(L\geq0\) and
\[
 t_i\in\{0,\ldots,n\},\qquad i=1,\ldots,M+L,\qquad
 \sum_i t_i=nL,
\]
\begin{equation}\label{ec:eq:rounded-cubic}
 \sum_i\psi_\theta(t_i)\geq0.
\end{equation}
Equivalently,
\[
 \sum_i\frac{t_i}{n}\left(1-\frac{t_i}{n}\right)
 \left(\bar a-\frac{t_i}{n}\right)\leq0.
\]
\end{lemma}

\begin{proof}
For a pair with sum \(s\),
\begin{equation}\label{ec:eq:pair-reduction}
 \psi_\theta(x)+\psi_\theta(s-x)
 =\psi_\theta(s)+[3s-2(n+\theta)]x(s-x).
\end{equation}
Choose a global minimizer and, among minimizers, one with as many boundary
coordinates as possible.  Let \(A_s=3s-2(n+\theta)\).  If \(A_s<0\), the
pairwise objective in \eqref{ec:eq:pair-reduction} is minimized by maximizing
\(x(s-x)\), so the two integer coordinates must differ by at most one.  If
\(A_s>0\), it is minimized by moving the pair to a feasible boundary.  If
\(A_s=0\), the objective is unchanged by such a move, which would produce
another minimizer with more boundary coordinates.  Hence every two
nonboundary coordinates in the selected minimizer differ by at most one.

Remove \(\ell\) coordinates equal to \(n\), which contribute zero, and write
\(h=L-\ell\).  If \(h=0\), the claim is immediate.  Otherwise let \(K\) be the
number of interior coordinates.  They sum to \(nh\), and
\[
 K\leq M+h,\qquad
 \mu=\frac{nh}{K}\geq\frac{nh}{M+h}\geq\frac{n}{M+1}=:\mu_0.
\]
The interior coordinates are the two integers bracketing \(\mu\).  Put
\[
 A_r=r(n-r),\qquad B_r=r^2(n-r),
\]
and linearly interpolate them on each interval \([r,r+1]\).  Then
\[
 \Theta_n(\mu)=\frac{B(\mu)}{A(\mu)}
\]
is nondecreasing because
\[
 \Theta_n'(\mu)=
 \frac{A_rB_{r+1}-B_rA_{r+1}}{A(\mu)^2}
 =\frac{A_rA_{r+1}}{A(\mu)^2}\geq0.
\]
Equation \eqref{ec:eq:theta} is exactly
\(\theta=\Theta_n(\mu_0)\).  Therefore the balanced interior block has
objective
\[
 K A(\mu)[\Theta_n(\mu)-\theta]\geq0.
\]
This proves the minimum, and hence every feasible vector, is nonnegative.
Allowing \(t_i=n\) only enlarges the feasible set relative to nonempty teams.
\end{proof}

\subsection{Exact first competitor, ties, and closed form}

\begin{proof}[Proof of Proposition~\ref{prop:quadratic-exact}, global-optimality part]
At \(q=\qGt\), sum the bound in
Lemma~\ref{ec:lem:supporting-cubic} over any feasible deficit vector.
Lemma~\ref{ec:lem:rounded-cubic} makes its right side nonpositive, so
\(\sum_i h_q(t_i)\leq0\).  Equations \eqref{ec:eq:h-condition} and
\eqref{ec:eq:deficit-duality} show that no partition beats the target.
The near-balanced competitor ties by construction and beats the target below
the crossing.  Strictness of the supporting and packing steps, together with
monotonicity in \(q\), gives uniqueness above it.
\end{proof}

The threshold equality set is exact.  Up to labels:

\begin{itemize}
\item If \(n\geq M+1\), the only maximizers at \(q=\qGt\) are the target and
the near-balanced partition with \(J-v\) teams of size \(n-u\) and \(v\)
teams of size \(n-u-1\).
\item If \(2\leq n\leq M\), the maximizers are
\[
 \pi_L=\bigl((n-1)^{\,nL},\,n^{\,M-(n-1)L}\bigr),
 \qquad
 0\leq L\leq\left\lfloor\frac{M}{n-1}\right\rfloor .
\]
Here \(h_q(0)=h_q(1)=0\) and
\(h_q(t)<0\) only for \(2\leq t\leq n-1\); the relaxed point \(t=n\)
always has \(h_q(n)=0\).
\end{itemize}
For \(n=2\), this ladder contains every feasible partition.

For a closed form, set \(m_0=J-v\), \(m_1=v\), and, for
\(t\in\{u,u+1\}\),
\[
 H_t(q)=t+\eta_tq,\quad \eta_t=t(2n-t-1),
\]
\[
 D_t(q)=(Mn-t)+\varepsilon_tq,\quad
 \varepsilon_t=Mn(n-1)-t(2n-t-1).
\]
Clearing the positive denominators in \eqref{ec:eq:qG2-tie} on \(q\geq0\)
gives
\[
 \Pi(q)=
 (M-1)\{m_0H_uD_{u+1}+m_1H_{u+1}D_u\}
 -D_uD_{u+1}
 =c_2q^2+c_1q+c_0,
\]
where
\begin{align*}
c_0={}&(M-1)[m_0u(Mn-u-1)+m_1(u+1)(Mn-u)]\\
&-(Mn-u)(Mn-u-1),\\
c_1={}&(M-1)\{m_0[u\varepsilon_{u+1}+\eta_u(Mn-u-1)]\\
&\quad+m_1[(u+1)\varepsilon_u+\eta_{u+1}(Mn-u)]\}\\
&-[(Mn-u)\varepsilon_{u+1}+\varepsilon_u(Mn-u-1)],\\
c_2={}&(M-1)[m_0\eta_u\varepsilon_{u+1}
 +m_1\eta_{u+1}\varepsilon_u]
-\varepsilon_u\varepsilon_{u+1}.
\end{align*}
The unique admissible root is stably evaluated as
\begin{equation}\label{ec:eq:qG2-closed}
 \qGt=\frac{-2c_0}{c_1+\sqrt{c_1^2-4c_2c_0}}.
\end{equation}
When \(u=0\) or \(v=0\), clearing denominators can introduce a negative
cancelled root.  Equation \eqref{ec:eq:qG2-closed} and all uniqueness claims
refer to the unique root on \(q\geq0\), not to every real root of the cleared
quadratic.

Two useful special cases are
\[
 (M+1)\mid n
 \quad\Longrightarrow\quad
 \qGt=\frac{1}{u(M^2-M-1)+1},
\]
\[
 n\leq M
 \quad\Longrightarrow\quad
 \qGt=\frac{1}{n(M-2)+2}.
\]
At the boundary \(n=M+1\), one has \(u=1\) and \(v=0\), and the first
displayed formula gives the same value \(1/[n(M-2)+2]\).  Thus the second
closed form remains valid for \(n\leq M+1\), although the statement
\(u=0\) does not.

\section{Quadratic stability--optimality separation}\label{ec:quadratic-gap}

\begin{proof}[Proof of Proposition~\ref{prop:quadratic-exact}, separation part]
Fix \(n>2\) and evaluate the near-balanced \((M+1)\)-team deficit vector at
\(q=\qBt\).  Delete its zero entries.  Its positive entries sum to \(n\), and
there are at least three of them: if \(u=0\), there are \(v=n>2\) unit
deficits; if \(u\geq1\), all \(M+1\geq3\) entries are positive.  Repeated use
of Lemma~\ref{ec:lem:superadd} gives
\[
 \sum_i g_{\qBt}(t_i)\leq g_{\qBt}(n)=1.
\]
At least the first merger combines two entries whose sum is strictly below
\(n\).  The lemma's equality characterization therefore makes the displayed
inequality strict.  By the deficit duality in
\eqref{ec:eq:deficit-duality}, the near-balanced competitor still has strictly
more aggregate effort than the target at \(q=\qBt\).

For every proper deficit \(t<n\), \(g_q(t)\) is strictly increasing in \(q\).
Hence the near-balanced deficit sum has a unique tie with one only at a
strictly larger value, so \(\qGt>\qBt\).  When \(n=2\), the near-balanced
vector is the canonical pair of unit deficits and its tie occurs at
\(1/[2(M-1)]\), giving equality of the two thresholds.
\end{proof}

For \(n>2\), the quadratic phase diagram is
\[
\begin{array}{ccl}
q\leq\qBt
&:&\text{stable, but not globally effort maximizing},\\[2pt]
\qBt<q<\qGt
&:&\text{unstable and not globally effort maximizing},\\[2pt]
q=\qGt
&:&\text{unstable and weakly globally effort maximizing},\\[2pt]
q>\qGt
&:&\text{unstable and uniquely globally effort maximizing}.
\end{array}
\]
For \(n=2\), the target is stable but not globally effort maximizing below
\(\qGt\), stable and nonuniquely globally effort maximizing at \(\qGt\), and
unstable and uniquely globally effort maximizing above \(\qGt\).
Thus stability never overlaps with \emph{strict} global optimality, including
the pair-team boundary case.

\section{Stable scale design}\label{ec:stable-scale}

\begin{proof}[Proof of Proposition~\ref{prop:stable-scale}]
The proof treats roster size as continuous only to establish monotonicity of
the exact frontier value, then restricts the result to population divisors.

For the linear benchmark, fix \(N\geq4\), let \(2\leq d\leq N/2\), and write
\[
 r=q_B(N/d,d)(d-1)>0.
\]
After substituting \(M=N/d\) in the defining equation for \(q_B\), the positive
root satisfies
\begin{equation}\label{ec:eq:scale-linear-root}
 H(N,d,r)
 =(N-d)(d-2)r^2+d(N-2d+2)r-d(d-1)=0.
\end{equation}
Moreover,
\begin{equation}\label{ec:eq:scale-linear-Hr}
 H_r=2(N-d)(d-2)r+d(N-2d+2)>0,
\end{equation}
including at \(d=2\).  The frontier effort is
\(V_L^S(d)=(N-d)(1+r)/N\).  Implicit differentiation gives
\begin{equation}\label{ec:eq:scale-linear-derivative}
 \frac{dV_L^S}{dd}
 =\frac{-(1+r)H_r-(N-d)H_d}{NH_r}.
\end{equation}
Equation~\eqref{ec:eq:scale-linear-root} is linear in \(N\) and yields
\begin{equation}\label{ec:eq:scale-linear-N}
 N=
 \frac{d\{(d-2)r^2+2(d-1)r+d-1\}}
 {r\{d+(d-2)r\}}.
\end{equation}
Substitution into the numerator of
\eqref{ec:eq:scale-linear-derivative} reduces it to
\begin{equation}\label{ec:eq:scale-linear-positive}
 -(1+r)H_r-(N-d)H_d
 =\frac{d\left(\{d+(d-2)r\}^2-2\right)}
 {\{d+(d-2)r\}^2}>0,
\end{equation}
because \(d+(d-2)r\geq d\geq2\).  Hence linear frontier effort is strictly
increasing on the continuous interval.

For the matched quadratic benchmark, put \(M=N/d\) and
\(r=q_B^{(2)}(M,d)(d-1)>0\).  Multiplying the threshold equation by \(d-1\)
gives
\begin{equation}\label{ec:eq:scale-quadratic-root}
 \Phi(M,d,r)=2d(M-1)^2r^2+B_2(M,d)r-(2M-1)(d-1)=0,
\end{equation}
where
\[
 B_2(M,d)=2M^2d-6Md+4M+3d-2.
\]
At the unique positive root,
\begin{equation}\label{ec:eq:scale-quadratic-phir}
 \Phi_r=
 \sqrt{B_2^2+8d(M-1)^2(2M-1)(d-1)}>0.
\end{equation}
Squared frontier effort is
\[
 W_Q^S(d)=\{V_Q^S(d)\}^2=d(M-1)(1+r).
\]
Along the fixed-population path \(N=Md\), differentiation uses
\(\mathcal D_N=\partial_d-(M/d)\partial_M\).  Applying this operator to
\eqref{ec:eq:scale-quadratic-root} and differentiating \(W_Q^S\) gives
\begin{equation}\label{ec:eq:scale-quadratic-derivative}
 \frac{dW_Q^S}{dd}
 =\frac{2Mdr+2Md-2Mr-3M-dr-d+1}{\Phi_r}.
\end{equation}
The numerator is increasing in \(M\), and \(M\geq2\) gives the strict lower
bound
\begin{equation}\label{ec:eq:scale-quadratic-positive}
 3d-5+(3d-4)r>0.
\end{equation}
Thus squared effort, and therefore effort, is strictly increasing in \(d\).

Restrict both continuous monotonicity results to the ordered divisor set
\(\mathcal D_N\).  For each fixed \(d\), target effort is strictly increasing
in \(q\), while Theorem~\ref{thm:linear-stability} and
Proposition~\ref{prop:quadratic-exact} identify the stable interval.  Its upper
endpoint is feasible under strict-block stability because the canonical
coalition is indifferent there.  Hence each roster selects its blocking
boundary, and the largest element of \(\mathcal D_N\) uniquely maximizes stable
effort.  If \(p\) is the smallest prime divisor of \(N\), that element is
\(N/p\), producing \(p\) equal teams.  For even \(N\), \(p=2\).
\end{proof}


\clearpage
\part*{Online Appendix}
\addcontentsline{toc}{part}{Online Appendix}
\noindent
This online appendix contains supporting extensions for the matched
convex-cost family, source span, retention, and high-curvature pair ordering.
Appendices A--G of the article contain the continuation mapping, the
split--clone proof, the full proofs of the exact linear and matched quadratic
results, and the stable-scale design result.  Sections OA.1--OA.2 restate the
comparative statics from Appendix D of the article at the isoelastic generality
used below.  Throughout, \(N=Mn\), the target
architecture is \(\target=(n,\ldots,n)\) with \(M\) full-cap teams, and
\(q\in[0,1]\) is the governance parameter.

\setcounter{section}{0}
\renewcommand{\thesection}{O\Alph{section}}
\renewcommand{\theHsection}{O\Alph{section}}

\section{Matched convex costs and deviation reduction}
\label{app:matched-costs}

\subsection{Isoelastic continuation and common intensity}

Let
\[
 c(x)=\frac{x^\rho}{\rho},\qquad \rho>1.
\]
For a team with effective value \(a_i>0\), the interior first-order condition
can be written
\begin{equation}\label{app:eq:iso-foc}
 a_i(1-p_i)=\lambda p_i^{\rho-1},
 \qquad \lambda=X^\rho .
\end{equation}
Because \(c'(0)=0\), every positive-value team is active.  For a fixed
\(\lambda/a_i\), the left and right sides of
\eqref{app:eq:iso-foc} determine a unique \(p_i\in(0,1)\).  Summing the
resulting probabilities determines a unique \(\lambda>0\).

Define
\[
 \varepsilon_i
 =
 \frac{p_i(1-p_i)}
 {(\rho-1)-(\rho-2)p_i}>0,
 \qquad
 \mathcal E=\sum_h\varepsilon_h.
\]
Log differentiation of \eqref{app:eq:iso-foc} gives
\[
 \left.\frac{\partial p_i}{\partial\log\lambda}\right|_{a_i}
 =-\varepsilon_i,
 \qquad
 \left.\frac{\partial p_i}{\partial\log a_i}\right|_{\lambda}
 =\varepsilon_i.
\]
If only \(a_j\) changes, probability adding-up implies
\[
 \frac{\partial\log\lambda}{\partial\log a_j}
 =\frac{\varepsilon_j}{\mathcal E}.
\]
Consequently,
\begin{align}
 \frac{\partial p_i}{\partial\log a_i}
 &=\varepsilon_i\left(1-\frac{\varepsilon_i}{\mathcal E}\right)>0,
 \label{app:eq:own-value}\\
 \frac{\partial p_i}{\partial\log a_j}
 &=-\frac{\varepsilon_i\varepsilon_j}{\mathcal E}<0,
 \qquad j\ne i.
 \label{app:eq:rival-value}
\end{align}
Raising a team's own effective value raises its probability; lowering any
rival value raises it as well.  Deleting a rival is obtained as its value
tends to zero.

\subsection{Payoff monotonicity and full-size completion}

Let total prize be \(P_k=Vk^\beta\), \(V>0\), and let the role-lottery value be
\[
 a_k=Vk^{\beta-1}(1-q+qk).
\]
Using \eqref{app:eq:iso-foc}, total effort cost of a size-\(k\) team with
probability \(p\) is \(a_kp(1-p)/\rho\).  Its per-member payoff is therefore
\begin{equation}\label{app:eq:Uk}
 U_k(p)
 =
 \frac{P_kp-a_kp(1-p)/\rho}{k}
 =
 Vk^{\beta-1}p
 \left[1-\frac{r_k}{\rho}(1-p)\right],
 \qquad
 r_k=q+\frac{1-q}{k}.
\end{equation}
For every \(k\),
\begin{equation}\label{app:eq:Uk-p}
 \frac{\partial U_k}{\partial p}
 =
 Vk^{\beta-1}
 \left[
 1-\frac{r_k}{\rho}
 +\frac{2r_kp}{\rho}
 \right]>0.
\end{equation}

For fixed \(p\), put \(h=(1-p)/\rho\in[0,1)\).  Apart from the positive
factor \(Vp\),
\[
 U_k(p)
 =
 Vp\left[(1-qh)k^{\beta-1}-(1-q)hk^{\beta-2}\right].
\]
The continuous derivative has the sign of
\begin{equation}\label{app:eq:size-bracket}
 (\beta-1)(1-qh)k-(\beta-2)(1-q)h.
\end{equation}
For \(1\leq\beta\leq2\), both terms make
\eqref{app:eq:size-bracket} nonnegative.  For \(\beta\geq2\), use \(k\geq1\)
and
\[
 (\beta-1)q+(\beta-2)(1-q)=\beta-2+q\leq\beta-1
\]
to bound it below by \((\beta-1)(1-h)>0\).  Hence
\begin{equation}\label{app:eq:size-monotonicity}
 U_{k+1}(p)\geq U_k(p)
 \qquad\text{for all }\beta\geq1.
\end{equation}
The effective value \(a_k\) is also nondecreasing in \(k\).

\begin{theorem}[Full-size completion]\label{app:thm:completion}
Suppose \(\rho>1\), \(\beta\geq1\), contestants are homogeneous, and a
coalition forms one cross-source team of size \(c<n\).  There exists a
deviation using the same incumbent sources and no larger source span that
forms a size-\(n\) replacement and gives every replacement member a payoff
weakly above that in the size-\(c\) continuation.  Thus any strictly blocking
smaller cross-source deviation implies a strictly blocking full-size
replacement.
\end{theorem}

\begin{proof}
Because \(c<n\), at least one member remains outside the coalition in a source
already used.  Add such a member.  The focal team grows from \(c\) to \(c+1\),
so its value weakly rises; the corresponding residual loses a member, so that
rival's value weakly falls or the rival disappears.  Equations
\eqref{app:eq:own-value}--\eqref{app:eq:rival-value} imply that the focal
probability weakly rises.  Equations \eqref{app:eq:Uk-p} and
\eqref{app:eq:size-monotonicity} imply that the payoff of a member in the
enlarged team is no lower.  Repeat until size \(n\) is reached.  All members of
the completed team receive the same ex ante payoff; if the initial coalition
strictly gained relative to the symmetric target payoff, both its original and
new members strictly gain after completion.
\end{proof}

\begin{proposition}[Convex-cost one-source protection]
\label{app:prop:one-source-convex}
For \(\rho>1\) and \(\beta\geq1\), the full-cap symmetric target is not
strictly blocked by a one-source split.  Consequently, stability against all
one-new-team deviations is equivalent to stability against all full-size
cross-source replacements.
\end{proposition}

\begin{proof}
A one-source split forming a size-\(c<n\) team leaves \(M-1\) untouched
full-size teams and a positive residual.  Since \(a_n\geq a_c\), each untouched
team has probability at least that of the new team.  Hence
\[
 1=(M-1)p_n+p_c+p_{n-c}>Mp_c,
\]
so \(p_c<1/M\).  Payoff monotonicity in probability and team size gives
\[
 U_c(p_c)<U_c(1/M)\leq U_n(1/M).
\]
The equivalence then follows from Theorem~\ref{app:thm:completion}.
\end{proof}

\subsection{The residual-score dictionary}

Return to the matched locus \(\beta=\rho\).  Let \(A=a_n\), and consider a
residual team of size \(k\) at the target intensity.  Define \(z_k\in(0,M)\)
so that its implied probability is \(z_k/M\).  Dividing its first-order
condition by the target team's condition yields
\begin{equation}\label{app:eq:score-dictionary}
 \frac{a_k}{A}
 =
 \frac{(M-1)z_k^{\rho-1}}{M-z_k}.
\end{equation}
The right side is strictly increasing in \(z_k\).  It follows immediately that
the target-intensity probability ordering and the effective-score ordering
coincide.

Suppose a full-size replacement extracts from \(\ell\) sources and leaves
positive residual sizes \(k_1,\ldots,k_\ell\).  At the target intensity, the
clone and \(M-\ell\) untouched target teams contribute
\((M-\ell+1)/M\) probability.  The residuals contribute
\(\sum_i z_{k_i}/M\).  Because the probability sum is decreasing in intensity,
the deviation lowers intensity, and hence strictly blocks by Split--Clone, if
and only if
\begin{equation}\label{app:eq:residual-sum}
 \sum_{i=1}^{\ell}z_{k_i}<\ell-1.
\end{equation}
Writing extraction shares \(u_i=(n-k_i)/n\) and
\[
 g_\rho(u)=1-z_{n(1-u)}
\]
turns \eqref{app:eq:residual-sum} into
\begin{equation}\label{app:eq:loss-sum}
 \sum_{i=1}^{\ell}g_\rho(u_i)>1,
 \qquad
 u_i>0,\quad \sum_i u_i=1.
\end{equation}

\subsection{Low-extraction convexity}

For a continuous residual share \(x=k/n\), normalized raid intensity
\[
 \delta=\frac{nq}{1+(n-1)q},
\]
and \(r=\rho-1\), equation \eqref{app:eq:score-dictionary} becomes
\begin{equation}\label{app:eq:z-continuous}
 \frac{(M-1)z(x)^r}{M-z(x)}
 =
 x^r[1-\delta(1-x)].
\end{equation}
Let
\[
 a(x)=\frac{\delta x}{1-\delta(1-x)},
 \qquad
 b(x)=\frac{z(x)}{M-z(x)}.
\]
Log differentiation of \eqref{app:eq:z-continuous} gives
\begin{equation}\label{app:eq:z-curvature}
 \operatorname{sgn}z''(x)
 =
 -\operatorname{sgn}
 \left\{
 \frac{r+a(x)^2}{[r+a(x)]^2}
 +
 \frac{b(x)^2-r}{[r+b(x)]^2}
 \right\}.
\end{equation}

\begin{lemma}[Concavity above population share]
\label{app:lem:concavity-share}
If \(x\geq1/2\) and \(z(x)\geq x\), then \(z''(x)<0\).
\end{lemma}

\begin{proof}
Put \(m=M-1\).  Equation \eqref{app:eq:z-continuous} and \(z\geq x\)
imply
\[
 a(x)\leq\frac{x}{m},
 \qquad
 b(x)\geq\frac{x}{m+1-x}.
\]
The first fraction inside braces in \eqref{app:eq:z-curvature} decreases in
\(a\in[0,1]\), whereas the second increases in \(b\).  It therefore suffices
to evaluate the brace at these two bounds.  With \(y=2x-1\in[0,1]\), its
numerator has the cubic Bernstein representation
\[
 \sum_{j=0}^{3}\binom{3}{j}B_jy^j(1-y)^{3-j},
\]
where
\begin{align*}
B_0&=4mr+r^2+r+2,\\
B_1&=\frac23(8m^2r^2+4mr^2+14mr+r^2+3r+6),\\
B_2&=\frac83(mr+1)(4mr+r+3),\\
B_3&=16(mr+1)^2.
\end{align*}
Every coefficient is strictly positive, so the brace is positive and
\eqref{app:eq:z-curvature} gives \(z''(x)<0\).
\end{proof}

Let \(q^{1/2}_{M,n}=1/[n(M-3/2)+1]\), the static score-equality boundary
between the target and a half-size residual.  The canonical matched threshold
satisfies
\begin{equation}\label{app:eq:midpoint-separation}
 q_{\rho;M,n}^{B,\mathrm{can}}<q^{1/2}_{M,n}
 \qquad (n>2).
\end{equation}
Lemma~\ref{app:lem:midpoint-separation} below proves this strict separation
by comparing the high-curvature limiting canonical equation at
\(q^{1/2}_{M,n}\) and then using global curvature monotonicity.

At and below the canonical onset, \eqref{app:eq:midpoint-separation} implies
\(z(x)>x\) for \(x\in[1/2,1)\).  Lemma
\ref{app:lem:concavity-share} then implies that
\[
 g_\rho(u)=1-z(1-u)
\]
is strictly increasing and strictly convex on \(u\in[0,1/2]\), with
\[
 g_\rho(0)=0,
 \qquad
 0<g_\rho(u)<u.
\]
In particular, for \(a,b\geq0\) and \(a+b\leq1/2\),
\begin{equation}\label{app:eq:low-superadd}
 g_\rho(a)+g_\rho(b)\leq g_\rho(a+b).
\end{equation}

\begin{theorem}[Matched-cost two-source sufficiency]
\label{app:thm:two-source}
Along \(\beta=\rho>1\), the first strictly blocking one-new-team deviation is a
full-size two-source replacement.  If \(q_t^B(\rho)\) is the threshold for
extraction vector \((t,n-t)\), then
\[
 q_{\rho;M,n}^{B,\mathrm{all}}
 =
 \min_{1\leq t\leq\lfloor n/2\rfloor}q_t^B(\rho).
\]
\end{theorem}

\begin{proof}
Full-size completion and one-source protection leave only full-size
cross-source compositions.  If \(n=2\), the only nontrivial full-size
cross-source composition is \((1,1)\), so the conclusion is immediate.
Henceforth suppose \(n>2\).  Because the canonical pair itself is indifferent
at \(q_{\rho;M,n}^{B,\mathrm{can}}\), the first onset cannot lie above that
value.  It therefore suffices to compare compositions on the parameter region
where the low-extraction convexity established above applies.  Consider shares
\(u_1,\ldots,u_\ell\) with
\(\sum_i u_i=1\).  If every share is at most \(1/2\), convexity implies that
the maximum of \(\sum_i g_\rho(u_i)\) over the capped simplex is attained at
an extreme point with two shares equal to \(1/2\); hence
\[
 \sum_i g_\rho(u_i)\leq2g_\rho(1/2)<1.
\]
Such a composition cannot be the first block.

Otherwise, exactly one share \(U\) exceeds \(1/2\), and the remaining shares
sum to \(s=1-U<1/2\).  Convexity, \(g_\rho(0)=0\), and
\eqref{app:eq:low-superadd} give
\[
 \sum_{i:u_i\ne U}g_\rho(u_i)\leq g_\rho(s).
\]
Thus the composition's loss is no larger than
\(g_\rho(U)+g_\rho(1-U)\), the loss from a two-source split.  The exact onset
is therefore the minimum over the finite integer pair list.
\end{proof}

The reduction also identifies several canonical-first regions.  The remaining
ordering problem is finite, with \(t=1,\ldots,\lfloor n/2\rfloor\).  The
canonical pair \(t=1\) is uniquely first for all \(n\leq4\), in a right
neighborhood of \(\rho=1\), in an open neighborhood of \(\rho=2\), and for
all sufficiently high \(\rho\), where each of these last three statements is
for a fixed finite \((M,n)\).
Theorem~\ref{app:thm:small-team-canonical} and
Proposition~\ref{app:prop:canonical-openness} prove the first three regions;
Appendix~\ref{app:high-curvature} proves the last.

\subsection{Global curvature monotonicity of the canonical threshold}

This subsection proves the compensated comparison used in
Section~\ref{app:matched-costs}.  Let \(r=\rho-1\), \(n>2\), and at the
canonical threshold write the normalized target-intensity probabilities of
the size-\((n-1)\) and singleton residuals as \(1-w\) and \(w\).  The
score--probability dictionary implies \(0<w<1/n\).  Put
\[
 D=1+q(n-1),\qquad B=1+q(n-2).
\]
The two residual equations and the indifference condition give
\begin{align}
 D(w,r)
 &=
 \frac{M-w}{M-1}(nw)^{-r},
 \label{app:eq:Dwr}\\
 B(w,r)
 &=
 \frac{M-w}{M-1+w}
 \left[\frac{1-w}{(n-1)w}\right]^r,
 \label{app:eq:Bwr}\\
 R(w,r)
 &:=(n-1)B(w,r)-(n-2)D(w,r)-1=0.
 \label{app:eq:Rwr}
\end{align}
The canonical threshold is
\begin{equation}\label{app:eq:qwr}
 q_{\rho;M,n}^{B,\mathrm{can}}
 =
 \frac{D(w(r),r)-1}{n-1}.
\end{equation}

Define
\[
 a=\log\frac1{nw},
 \qquad
 c=\log\frac{n(1-w)}{n-1},
 \qquad
 b=a+c,
\]
and
\[
 d=\frac1{M-w}+\frac rw,
 \qquad
 e=d+\frac1{M-1+w}+\frac r{1-w}.
\]
Then \(D_r=aD\), \(D_w=-dD\), \(B_r=bB\), and \(B_w=-eB\).
At the threshold put \(U=(n-2)D\), so \((n-1)B=U+1\).  It follows that
\[
 R_r=Uc+b>0,
 \qquad
 R_w=-e-U(e-d)<0,
\]
and hence \(dw/dr>0\).

Differentiating \eqref{app:eq:qwr} and using the implicit derivative yields
\begin{equation}\label{app:eq:D-derivative}
 \frac1D\frac{dD}{dr}
 =
 \frac{(U+1)(ae-bd)}{e+U(e-d)}.
\end{equation}
The denominator is positive.  Since \(b=a+c\),
\begin{equation}\label{app:eq:ae-bd}
 ae-bd
 =
 \frac a{M-1+w}-\frac c{M-w}
 +r\left[\frac a{1-w}-\frac cw\right].
\end{equation}
The Cauchy mean-value theorem gives
\[
 \frac ac
 \in
 \left(n-1,\frac{1-w}{w}\right).
\]
Thus the constant term in \eqref{app:eq:ae-bd} is positive and its
coefficient on \(r\) is negative.  The unique sign-change value is
\begin{equation}\label{app:eq:rstar}
 r^*(w)
 =
 \frac{
 \dfrac a{M-1+w}-\dfrac c{M-w}}
 {\dfrac cw-\dfrac a{1-w}}>0.
\end{equation}
It remains to show that the threshold root lies below \(r^*(w)\).

Divide \eqref{app:eq:Rwr} by \(D\).  With
\[
 K_0=\frac{M-1}{M-1+w},
 \qquad
 t_0=\frac{M-1}{M-w},
\]
the threshold equation becomes
\[
 (n-1)K_0e^{rc}=n-2+t_0e^{-ra}.
\]
Since \(a,r>0\),
\begin{equation}\label{app:eq:r-upper}
 r<\frac{\log R_0}{c},
 \qquad
 R_0=\frac{n-2+t_0}{(n-1)K_0}.
\end{equation}

\begin{lemma}[Weighted-log bound]\label{app:lem:weighted-log}
For \(M\geq2\), \(n\geq3\), and \(0<w<1/n\), whenever \(R_0>1\),
\[
 \frac{\log R_0}{c}<r^*(w).
\]
\end{lemma}

\begin{proof}
Let \(t=a/c\) and
\[
 F(t)
 =
 \frac{\dfrac t{M-1+w}-\dfrac1{M-w}}
 {\dfrac1w-\dfrac t{1-w}}.
\]
Direct differentiation gives
\[
 F'(t)
 =
 \frac{Mw(1-w)(1-2w)}
 {(M-1+w)(M-w)[1-w-tw]^2}>0.
\]
Since \(a/c>n-1\),
\[
 cr^*(w)=cF(a/c)>cF(n-1).
\]
The inequality \(\log(1+x)\geq x/(1+x)\) gives
\[
 cF(n-1)
 \geq
 \underline L(w)
 =
 \frac{w[(n-2)M+1-nw]}
 {n(M-w)(M-1+w)}.
\]
It remains to compare \(\underline L\) with \(R_0-1\).  Put
\[
 m=M-2,\qquad s=n-3,\qquad x=1-nw.
\]
After multiplication by positive denominators,
\(\underline L-(R_0-1)\) has numerator
\[
 P(m,s,x)=C_0+C_1x-C_2x^2-(s+1)x^3,
\]
where
\begin{align*}
C_0={}&m^2s^3+6m^2s^2+11m^2s+6m^2
 +2ms^3+12ms^2+22ms+12m\\
&+s^3+5s^2+8s+4,\\
C_1={}&3m^2s^2+16m^2s+21m^2
 +7ms^2+39ms+54m+6s^2+28s+36,\\
C_2={}&ms^2+7ms+12m+2s^2+8s+12.
\end{align*}
For \(0<x<1\), \(x^2\leq x\) and \(x^3\leq x\).  Hence
\[
 P\geq C_0+[C_1-C_2-(s+1)]x>0,
\]
because
\[
 C_1-C_2-(s+1)
 =
 3m^2s^2+16m^2s+21m^2
 +6ms^2+32ms+42m+4s^2+19s+23.
\]
Therefore
\[
 \log R_0<R_0-1\leq\underline L<cr^*(w),
\]
which proves the claim.
\end{proof}

\begin{theorem}[Global curvature monotonicity]
\label{app:thm:curvature-monotonicity}
Along the compensated matched locus \(\beta=\rho\),
\[
 q_{\rho;M,2}^{B,\mathrm{can}}=\frac1{2(M-1)}
\]
for every \(\rho>1\).  If \(n>2\), then
\[
 \frac{d q_{\rho;M,n}^{B,\mathrm{can}}}{d\rho}>0.
\]
Moreover,
\[
 q_{1;M,n}^{B,\mathrm{can}}
 <
 q_{\rho;M,n}^{B,\mathrm{can}}
 <
 q_{\infty;M,n}^{B,\mathrm{can}}
 <
 \frac1{n(M-2)+2}.
\]
\end{theorem}

\begin{proof}
Equations \eqref{app:eq:r-upper} and
Lemma~\ref{app:lem:weighted-log} give \(r<r^*(w)\).  By
\eqref{app:eq:ae-bd}, \(ae-bd>0\); equation
\eqref{app:eq:D-derivative} then implies \(dD/dr>0\), and
\eqref{app:eq:qwr} gives the strict derivative.  For \(n=2\), the residual
pair is balanced and the equation reduces directly to \(1/[2(M-1)]\).
The remaining upper chain follows from
Lemma~\ref{app:lem:midpoint-separation} below.  That lemma uses only the strict
curvature monotonicity established above and the high-curvature limiting
equation.
\end{proof}

\subsection{Exact small-team and benchmark-neighborhood ordering}

\begin{lemma}[Midpoint separation]\label{app:lem:midpoint-separation}
For \(M\geq2\), \(n>2\), and every finite \(\rho>1\) on the matched locus,
\[
 q_{\rho;M,n}^{B,\mathrm{can}}<q^{1/2}_{M,n}.
\]
\end{lemma}

\begin{proof}
Let \(q_{\infty;M,n}^{B,\mathrm{can}}\) be the high-curvature limit in
Theorem~\ref{app:thm:curvature-monotonicity}.  Taking the limit in the two
canonical residual-score equations and eliminating the limiting residual
probabilities gives the scalar equation
\begin{equation}\label{app:eq:canonical-limit-L}
 \mathcal L_{M,n}(q)
 :=
 \frac{[1+q(n-2)]^{n-1}}{[1+q(n-1)]^n}
 \frac{[n(M-1)+1]^{n-1}(nM-1)}
 {[n(M-1)]^n}
 =1.
\end{equation}
The logarithmic derivative is
\[
 \frac{d\log\mathcal L_{M,n}(q)}{dq}
 =
 (n-1)\left\{
 \frac{n-2}{1+q(n-2)}
 -
 \frac{n}{1+q(n-1)}
 \right\}<0.
\]
Thus \(\mathcal L_{M,n}\) crosses one from above to below at the limiting
canonical threshold.

Put \(s=n(M-1)\) and
\[
 \xi=\frac{n-2}{2s(s+n/2)}>0.
\]
Direct substitution of \(q^{1/2}_{M,n}=1/[n(M-3/2)+1]\) into
\eqref{app:eq:canonical-limit-L} yields
\begin{equation}\label{app:eq:L-midpoint-factor}
 \mathcal L_{M,n}\!\left(q^{1/2}_{M,n}\right)
 =
 [1-(n-1)\xi](1+\xi)^{n-1}.
\end{equation}
Here \(0<(n-1)\xi<1\).  Moreover,
\[
 \frac{d}{d\xi}
 \left\{
 \log[1-(n-1)\xi]+(n-1)\log(1+\xi)
 \right\}
 =
 -\frac{n(n-1)\xi}
 {[1-(n-1)\xi](1+\xi)}<0.
\]
The expression in braces is zero at \(\xi=0\), so
\eqref{app:eq:L-midpoint-factor} is strictly below one.  Since
\(\mathcal L_{M,n}\) is strictly decreasing,
\[
 q_{\infty;M,n}^{B,\mathrm{can}}<q^{1/2}_{M,n}.
\]
The strict curvature monotonicity established above places every
finite-curvature canonical threshold below its limiting value, proving the
result.
\end{proof}

\begin{theorem}[Canonical-first for small teams]
\label{app:thm:small-team-canonical}
Fix \(M\geq2\) and let \(n\in\{2,3,4\}\).  Along the matched locus
\(\beta=\rho>1\), the canonical extraction pair \((1,n-1)\) is, up to source
labels, the unique first blocking pair.  Together with
Theorem~\ref{app:thm:two-source}, it therefore determines the exact complete
one-new-team stability threshold.
\end{theorem}

\begin{proof}
For \(n=2\), the only feasible two-source extraction is \((1,1)\); for
\(n=3\), the only pair up to labels is \((1,2)\).  Let \(n=4\).  The only
pairs are the canonical pair \((1,3)\) and the balanced pair \((2,2)\).
The balanced pair is indifferent exactly when
\[
 2g_\rho(1/2)=1
 \quad\Longleftrightarrow\quad
 z(1/2)=1/2.
\]
Substituting \(x=z=1/2\) into the residual-score dictionary and cancelling
the common factor \(x^{\rho-1}\) gives
\[
 \frac{1+q(n/2-1)}{1+q(n-1)}
 =
 \frac{M-1}{M-1/2},
\]
whose unique solution is \(q=q^{1/2}_{M,n}\).  For every proper residual
size \(k<n\), the ratio \(a_k/a_n\) strictly decreases in \(q\).
Equation~\eqref{app:eq:score-dictionary} therefore implies that \(z_k\)
strictly decreases and \(g_\rho\) strictly increases in \(q\).  Hence the
balanced pair's threshold is exactly \(q^{1/2}_{M,n}\).  By
Lemma~\ref{app:lem:midpoint-separation}, the canonical threshold is strictly
lower.  Matched two-source sufficiency excludes every higher-source
composition from preceding these pairs.
\end{proof}

\begin{proposition}[Canonical-first near the exact benchmarks]
\label{app:prop:canonical-openness}
For every fixed \(M,n\geq2\), there are constants
\(\varepsilon_1(M,n)>0\) and \(\varepsilon_2(M,n)>0\) such that the canonical
pair is uniquely first whenever
\[
 1<\rho<1+\varepsilon_1(M,n)
 \quad\text{or}\quad
 |\rho-2|<\varepsilon_2(M,n)
\]
on the matched locus.
\end{proposition}

\begin{proof}
For \(t=1,\ldots,\lfloor n/2\rfloor\), define the pair margin
\[
 \Phi_t(q,\rho)
 =
 g_\rho(t/n)+g_\rho(1-t/n)-1.
\]
The residual-score dictionary makes every \(\Phi_t\) continuous in
\((q,\rho)\), including its limit at \(\rho=1\), and strictly increasing in
\(q\).  The latter follows because \(a_k/a_n\) strictly decreases in \(q\)
for every \(k<n\), while the right side of
\eqref{app:eq:score-dictionary} strictly increases in \(z_k\).
In particular, \(\partial_q\Phi_1>0\), so the implicit-function theorem gives
a locally unique continuous canonical root \(q_1^B(\rho)\).

At \(\rho=1\), the strict-increment argument in the canonical-envelope lemma
of Appendix C of the article gives
\[
 \Phi_t(q_1^B(1),1)<0
 \qquad\text{for every }t>1.
\]
At \(\rho=2\), the quadratic residual-loss argument in Appendix D of the
article gives the same strict inequalities at \(q_1^B(2)=\qBt\).  There are
only finitely many noncanonical pairs.
Continuity therefore preserves all of these strict gaps in a right
neighborhood of \(1\) and an open neighborhood of \(2\).  Since each
\(\Phi_t\) rises with \(q\), no noncanonical pair can cross zero before the
canonical root in either neighborhood.  Theorem~\ref{app:thm:two-source}
then extends the ordering from pairs to the complete one-new-team class.
\end{proof}

The derivative is a compensated comparison along the matched locus, where
prize and cost curvature rise together.

\section{Finite packing, source span, and retention}
\label{app:source-span}\label{ec:scope}

The proofs separate onset from severity.  In both exact benchmarks, two-source
canonical replacement determines the first loss of one-new-team stability.
This does not imply that it maximizes the coalition gain at every larger
\(q\); multi-source replacements can become more damaging later.  A cap
restricts final team size, whereas a source-span rule restricts residual-team
geometry.  One-source recruitment is protected, while allowing two sources
admits the canonical block.

For the finite formulas below, \(R\) is an effective integer source cap and
\[
 2\leq R\leq\min\{M,n\}.
\]
A nominal cap larger than this range is replaced by
\(R_{\mathrm{eff}}=\min\{R,M,n\}\).  The case \(R=1\) permits only a
protected smaller split or a full-team relabeling; by convention its required
retention friction is zero.  The continuous relaxation can use
\(2\leq R\leq M\), while every exact finite envelope, plateau, and retention
statement uses the finite effective range above.

Retention charges or consent constraints can restore a target by changing the
formation payoff.  They do not change which static partition generates the
most effort.  The results below characterize how source-span restrictions and
retention frictions shape the severity of formation pressure beyond its onset.

\subsection{Continuous source-span envelope and its large-team meaning}

For a full-size linear replacement, let \(x_i=t_i/n\) be the extraction share
from incumbent \(i\) and
\[
 \delta=\frac{nq}{1+(n-1)q}.
\]
The finite model restricts \(x_i\) to the \(1/n\) grid.  Its continuous closure
is
\[
 x_i\geq0,\qquad
 \sum_{i=1}^{M}x_i=1,\qquad
 \#\{i:x_i>0\}\leq R.
\]
If the clone has probability \(p\), residual \(i\)'s probability is
\[
 \phi_p(\delta x_i)
 =
 \begin{cases}
 \dfrac{p-\delta x_i}{1-\delta x_i},
 &\delta x_i<p,\\[6pt]
 0,&\delta x_i\geq p.
 \end{cases}
\]
The probability equation is
\begin{equation}\label{app:eq:continuous-prob}
 p+\sum_{i=1}^{M}\phi_p(\delta x_i)=1.
\end{equation}
On \([0,p]\), \(\phi_p\) is strictly decreasing and strictly concave.
Holding total raid mass fixed, repeated transfers between interior coordinates
therefore weakly lower the rival-probability sum until all but at most one
positive coordinate lie at the boundary \(p\).  If \(r\) residuals are exactly
inactive and one is partially active, the raid-mass vector is
\[
 \underbrace{p,\ldots,p}_{r\text{ coordinates}},
 \quad \delta-rp,\quad 0,\ldots,0.
\]

The adding-up equation becomes
\[
 r(M-r)p^2+[1+(M-r)(1-\delta)]p-1=0.
\]
Define
\[
 b_r(\delta)=1+(M-r)(1-\delta)
\]
and
\begin{equation}\label{app:eq:pbar-r}
 \bar p_r(\delta)
 =
 \frac{2}{
 b_r(\delta)+
 \sqrt{b_r(\delta)^2+4r(M-r)}
 }.
\end{equation}
Adjacent regimes meet at
\begin{equation}\label{app:eq:delta-r}
 \delta_r=\frac{r}{M-r+1}.
\end{equation}
If the source cap binds and \(\delta\geq\delta_R\leq1\), all \(R\) residuals
can be made inactive and the envelope is \(1/(M-R+1)\).  Otherwise,
\begin{equation}\label{app:eq:continuous-r-index}
 r=
 \min\left\{
 R-1,
 \left\lfloor\frac{\delta(M+1)}{1+\delta}\right\rfloor
 \right\},
\qquad
 \bar p_{M,R}(\delta)=\bar p_r(\delta).
\end{equation}
At a boundary, adjacent formulas agree.

This is a limit theorem, not a finite-grid identity.  If
\(\delta_n(q_n)\to\delta\), density of the extraction grid and continuity of
the unique root in \eqref{app:eq:continuous-prob} imply
\[
 p^*_{M,n,R}(q_n)\longrightarrow\bar p_{M,R}(\delta).
\]
For each finite \(n\), the continuous value is an upper bound.  The exact
finite correction is derived below.

The continuous solution also gives a transparent source-span ladder.  Within
regime \(r\), the extraction shares are
\[
 \underbrace{\frac p\delta,\ldots,\frac p\delta}_{r\text{ shares}},
 \qquad
 1-\frac{rp}{\delta}.
\]
The final share enters at zero when \(\delta=\delta_r\) and reaches the common
share \(1/(r+1)\) at \(\delta_{r+1}\).  The corresponding relaxed governance
switch is
\begin{equation}\label{app:eq:qhat-r}
 \widehat q_{r,n}
 =
 \frac{r}{n(M+1-2r)+r}.
\end{equation}
Because \(\delta\leq1\), the largest source span with a genuine continuous
margin is
\[
 R_0=\left\lfloor\frac M2\right\rfloor+1.
\]
Source rights beyond \(R_0\) can improve only finite-grid rounding, and that
gain vanishes in the large-team limit.

At maximal governance, the unrestricted continuous envelope reduces to
\begin{equation}\label{app:eq:q1-envelope}
 \bar p_M(1)
 =
 \frac{2}{
 1+\sqrt{1+4\lfloor M^2/4\rfloor}
 }.
\end{equation}
The value is already attained once \(R\geq R_0\).  This special case
illustrates the difference between onset and saturation: two sources create
the first block, while up to roughly half of the incumbent field can matter
for the largest high-governance gain.

\subsection{Probability equation for a full-size raid}

In the linear benchmark, set
\[
 A=1+(n-1)q,\qquad h=\frac{q}{A}.
\]
Represent a full-size deviation by \(M\) extraction amounts
\(t_1,\ldots,t_M\), allowing zeros, with
\[
 0\leq t_i\leq n-1,\qquad
 \sum_{i=1}^{M}t_i=n,\qquad
 \#\{i:t_i>0\}\leq R.
\]
The new clone has value \(A\), and source \(i\)'s residual has value
\(A(1-ht_i)\).  If the clone's equilibrium probability is \(p\), aggregate
effort is \(A(1-p)\).  Residual \(i\) is active if and only if \(ht_i<p\);
when active, its probability is
\[
 \phi_p(ht_i)
 =
 \frac{p-ht_i}{1-ht_i}.
\]
Using \(\phi_p(y)=0\) for \(y\geq p\), the exact probability equation is
\begin{equation}\label{app:eq:finite-probability}
 p+\sum_{i=1}^{M}\phi_p(ht_i)=1.
\end{equation}
This formulation includes untouched incumbents as \(t_i=0\), for which
\(\phi_p(0)=p\).

\subsection{Integer compression by activity buckets}

For \(q>0\), assign a candidate probability to bucket \(s\in\{1,\ldots,n\}\).
If \(s<n\), the bucket condition is
\begin{equation}\label{app:eq:bucket}
 h(s-1)<p\leq hs;
\end{equation}
extractions \(t\leq s-1\) leave active residuals, while \(t\geq s\) leave
inactive residuals.  The all-active bucket is \(s=n\), with
\(p>h(n-1)\).  Write \(u=s-1\) and define, for active residuals,
\begin{equation}\label{app:eq:gh}
 g_h(t)=\frac{ht}{1-ht},
 \qquad
 \phi_p(ht)=p-(1-p)g_h(t).
\end{equation}
The function \(g_h\) is strictly increasing and strictly convex.

Let \(d\in\{0,\ldots,R\}\) be the number of inactive residuals.  They absorb
at least \(ds\) recruits; the remaining active sources are at most \(R-d\)
and each absorbs at most \(u\).  Define
\begin{equation}\label{app:eq:Sds}
 S_{d,s}
 =
 \min\{n-ds,(R-d)u\}.
\end{equation}
The pair \((d,s)\) is packing-feasible if
\begin{equation}\label{app:eq:packing-feasible}
 0\leq d\leq R,\qquad ds\leq n,\qquad
 S_{d,s}\geq\max\{0,n-d(n-1)\}.
\end{equation}
When \(u>0\), write
\[
 S_{d,s}=k_{d,s}u+a_{d,s},
 \qquad 0\leq a_{d,s}<u;
\]
for \(u=0\), set \(k_{d,s}=a_{d,s}=0\).  Convexity of \(g_h\) implies that
the maximum active-residual loss packs the extraction mass as
\[
 \underbrace{u,\ldots,u}_{k_{d,s}\text{ times}},
 \quad
 a_{d,s}\ \text{if positive},
 \quad
 0,\ldots,0.
\]
Accordingly define
\begin{equation}\label{app:eq:Gds}
 G_{d,s}(q)
 =
 k_{d,s}g_h(u)
 +\ones\{a_{d,s}>0\}g_h(a_{d,s}).
\end{equation}

\begin{lemma}[Finite raid-mass packing]\label{app:lem:raid-packing}
Fix an activity bucket \(s\) and an inactive count \(d\).  Among all integer
extraction vectors satisfying the source cap and
\eqref{app:eq:packing-feasible}, the packed vector above minimizes rival
probability in \eqref{app:eq:finite-probability}.
\end{lemma}

\begin{proof}
For active extractions \(0\leq x\leq y<u\), moving one unit from \(x\) to
\(y\) weakly raises \(g_h(x)+g_h(y)\), by discrete convexity.  Repeating the
transfer puts every positive active extraction except at most one at the upper
bound \(u\).  Inactive residuals jointly absorb \(n-S_{d,s}\); their internal
allocation does not enter \eqref{app:eq:finite-probability}.
\end{proof}

In the packed structure, the probability adding-up equation is
\[
 (M-d+1)p-(1-p)G_{d,s}=1.
\]
Its root is
\begin{equation}\label{app:eq:pds}
 p_{d,s}(q)
 =
 \frac{1+G_{d,s}(q)}{M-d+1+G_{d,s}(q)}.
\end{equation}
Let \(\mathcal C_{M,n,R}(q)\) contain the packing-feasible pairs whose roots
satisfy the bucket condition \eqref{app:eq:bucket}, or the all-active
condition when \(s=n\).

\begin{theorem}[Exact finite source-span envelope]
\label{app:thm:finite-envelope}
For \(M,n\geq2\), \(2\leq R\leq\min\{M,n\}\), and \(q>0\), the largest
winning probability attainable by a full-size one-new-team replacement is
\begin{equation}\label{app:eq:pstar}
 p^*_{M,n,R}(q)
 =
 \max_{(d,s)\in\mathcal C_{M,n,R}(q)}p_{d,s}(q).
\end{equation}
At \(q=0\), \(p^*_{M,n,R}(0)=1/(M+1)\).  At most \(n(R+1)\) candidates need
be evaluated.
\end{theorem}

\begin{proof}
For an extraction vector \(\mathbf t\), let
\[
 F_{\mathbf t}(p)
 =
 p+\sum_i\phi_p(ht_i)-1.
\]
Its equilibrium probability is the unique root.  Conditional on \(p\), Lemma
\ref{app:lem:raid-packing} exactly minimizes \(F_{\mathbf t}(p)\) inside each
bucket and inactive-count class.  The minimized expression is linear in \(p\)
and has root \eqref{app:eq:pds}.  Taking the largest self-consistent root
therefore gives the global integer envelope, not a relaxation.
\end{proof}

The theorem also explains the finite correction to the continuous
source-span compression.  In the continuum, active extraction can be
concentrated at a boundary and one partial source.  On the integer grid,
several active sources may be tied at \(s-1\), plus at most one remainder.
For example, an optimizer can have extraction vector \((3,3,4)\) when
\((M,n)=(5,10)\).

\subsection{Plateaus, switches, and finite jumps}

With source cap \(R\), at least \(M-R\) incumbents remain untouched.  Including
the clone, at least \(M-R+1\) equal value-\(A\) teams are active, so
\[
 p^*_{M,n,R}(q)\leq\frac{1}{M-R+1}.
\]
Equality requires exactly \(R\) sources and inactivity of every raided
residual.  Define
\begin{equation}\label{app:eq:Dplateau}
 D_{M,n,R}
 =
 (M-R+1)\left\lfloor\frac nR\right\rfloor-(n-1).
\end{equation}

\begin{theorem}[Exact finite cap plateau]\label{app:thm:plateau}
For \(M,n\geq2\) and \(2\leq R\leq\min\{M,n\}\), if
\(D_{M,n,R}\leq0\), the complete source-cap plateau is unreachable for
\(q\in[0,1]\).  If \(D_{M,n,R}>0\), its exact onset is
\[
 q^{\mathrm{plat}}_{M,n,R}=\frac1{D_{M,n,R}},
\]
and for every larger \(q\),
\[
 p^*_{M,n,R}(q)=\frac1{M-R+1}.
\]
\end{theorem}

\begin{proof}
At the plateau probability, each raided residual must satisfy
\[
 ht_i\geq\frac1{M-R+1}.
\]
There exist \(R\) integer extraction amounts summing to \(n\) and satisfying
this inequality if and only if its required lower bound does not exceed
\(\lfloor n/R\rfloor\).  Using \(1/h=1/q+n-1\) gives
\eqref{app:eq:Dplateau} and the stated threshold.
\end{proof}

The envelope is weakly increasing in \(q\), because \(h(q)\) rises and lowers
every residual value relative to the clone.  The source count of an optimal
integer vector need not rise one at a time.  For \(M=n=4\), \((2,2)\) is
optimal below \(q=1/2\); at \(q=1/2\), the vectors
\((2,2)\), \((1,1,2)\), and \((1,1,1,1)\) all give \(p=1/3\); above the tie,
\((1,1,1,1)\) is uniquely optimal.  The minimum optimal span jumps from two
to four.

Every finite switch can nevertheless be calculated exactly.  It is either an
activity boundary
\[
 p_{d,s}(q)=hs
 \quad\text{or}\quad
 p_{d,s}(q)=h(s-1),
\]
or a pairwise candidate tie.  If \(G_i=G_{d_i,s_i}\), the latter is
\begin{equation}\label{app:eq:candidate-tie}
 (d_i-d_j)+(M-d_j)G_i-(M-d_i)G_j=0.
\end{equation}
Clearing the positive \(1-ht\) denominators makes each switch a finite
polynomial equation.

\subsection{Exact uniform retention friction}

For a full-size team in the linear benchmark,
\[
 U(p,r_n)=p[1-r_n(1-p)],
 \qquad
 r_n(q)=\frac{1+(n-1)q}{n}.
\]
By the full-size reduction underlying Theorem 2 of the article, the
probability envelope in \eqref{app:eq:pstar} also identifies the largest
one-new-team formation payoff.

\begin{proposition}[Exact retention schedule]\label{app:prop:retention}
For \(M,n\geq2\) and \(2\leq R\leq\min\{M,n\}\), the smallest uniform
per-member friction that deters every admissible one-new-team deviation under
source cap \(R\) is
\begin{equation}\label{app:eq:tau-star}
 \tau^*_{M,n,R}(q)
 =
 \left[
 U\!\left(p^*_{M,n,R}(q),r_n(q)\right)
 -
 U\!\left(\frac1M,r_n(q)\right)
 \right]_+.
\end{equation}
Equivalently,
\[
 \tau^*_{M,n,R}(q)
 =
 \left[
 \left(p^*-\frac1M\right)
 \left\{1-r_n+r_n\left(p^*+\frac1M\right)\right\}
 \right]_+.
\]
\end{proposition}

For \(R\geq2\), the friction is zero exactly when \(q\leq\qB\).  In the
all-active region, the canonical coalition attains the envelope.  At larger
\(q\), the canonical gain is only a lower bound when a different packed vector
attains \(p^*\).

On a profitable plateau, \(p^*\) is constant while \(r_n\) continues to rise.
Direct differentiation gives
\begin{equation}\label{app:eq:tau-plateau}
 \frac{d\tau^*}{dq}
 =
 -\frac{n-1}{n}
 \left(p^*-\frac1M\right)
 \left(1-p^*-\frac1M\right)<0.
\end{equation}
Formation pressure can therefore be nonmonotone in governance: once source
span binds the clone's probability, further concentration increases its
members' expected effort exposure rather than weakening rivals further.

\section{High-curvature pair ordering}
\label{app:high-curvature}

\subsection{The limiting two-source equation}

Along the matched locus, put \(r=\rho-1\) and let
\(x\in(0,1/2]\) denote the smaller residual share in a two-source replacement.
At normalized raid intensity \(\delta\), equation
\eqref{app:eq:z-continuous} is
\[
 \frac{(M-1)z_r(x;\delta)^r}{M-z_r(x;\delta)}
 =
 x^r[1-\delta(1-x)].
\]
The pair is indifferent when
\begin{equation}\label{app:eq:Sr}
 S_r(x,\delta)
 :=
 z_r(x;\delta)+z_r(1-x;\delta)-1=0.
\end{equation}
For fixed \(x\), \(S_r\) strictly decreases in \(\delta\), so its root is
unique.

Taking logs gives
\[
 r\log\frac{z_r(x;\delta)}x
 =
 \log[1-\delta(1-x)]
 +\log[M-z_r(x;\delta)]
 -\log(M-1).
\]
Define
\begin{equation}\label{app:eq:ell}
 \ell_\delta(x)
 =
 \log
 \frac{[1-\delta(1-x)](M-x)}{M-1}.
\end{equation}
Uniformly on compact subsets of \(x\in(0,1)\) and admissible \(\delta\),
\begin{equation}\label{app:eq:z-expansion}
 z_r(x;\delta)
 =
 x+\frac{x}{r}\ell_\delta(x)+O(r^{-2}).
\end{equation}
Hence
\begin{equation}\label{app:eq:S-expansion}
 S_r(x,\delta)
 =
 \frac1rJ_\delta(x)+O(r^{-2}),
 \qquad
 J_\delta(x)
 =
 x\ell_\delta(x)+(1-x)\ell_\delta(1-x).
\end{equation}
Moreover,
\[
 \frac{\partial J_\delta(x)}{\partial\delta}
 =
 -x(1-x)
 \left[
 \frac1{1-\delta(1-x)}
 +\frac1{1-\delta x}
 \right]<0.
\]
The high-curvature pair threshold \(\delta_\infty^B(x)\) is therefore the
unique root of \(J_\delta(x)=0\).

\subsection{A one-crossing representation}

Introduce
\[
 y=1-2x,\qquad
 \beta_M=\frac1{2M-1},\qquad
 \alpha=\frac{\delta}{2-\delta}.
\]
Then \(J_\delta(x)=F_M(y,\alpha)\), where
\begin{align}
F_M(y,\alpha)
={}&
\frac12\log[(1-\alpha^2y^2)(1-\beta_M^2y^2)]
\notag\\
&+y[\operatorname{arctanh}(\alpha y)
 -\operatorname{arctanh}(\beta_M y)]
\notag\\
&-\log[(1+\alpha)(1-\beta_M)].
\label{app:eq:F}
\end{align}
The parameter derivative is
\begin{equation}\label{app:eq:F-alpha}
 F_{M,\alpha}(y,\alpha)
 =
 -\frac{1-y^2}{(1+\alpha)(1-\alpha^2y^2)}<0.
\end{equation}
Useful endpoint identities are
\[
 F_M(1,\alpha)=0,
 \qquad
 F_M(y,\beta_M)
 =
 \log\frac{1-\beta_M^2y^2}{1-\beta_M^2}>0,
\]
and, for
\[
 \alpha_H=\frac{\beta_M}{1-\beta_M}=\frac1{2(M-1)},
\]
\[
 F_M(0,\alpha_H)=F_M(1,\alpha_H)=0.
\]

The sign of the \(y\)-derivative is controlled by
\begin{equation}\label{app:eq:F-series}
 \frac{F_{M,y}(y,\alpha)}y
 =
 \sum_{k=0}^{\infty}c_k(\alpha,\beta_M)y^{2k},
\end{equation}
with
\begin{equation}\label{app:eq:ck}
 c_k
 =
 \alpha^{2k+1}
 \left(1-\alpha+\frac1{2k+1}\right)
 -
 \beta_M^{2k+1}
 \left(1+\beta_M+\frac1{2k+1}\right).
\end{equation}
For \(p=2k+1\), the sign of \(c_k\) is the sign of
\[
 L(p)
 =
 p\log\frac{\alpha}{\beta_M}
 +\log\left(1-\alpha+\frac1p\right)
 -\log\left(1+\beta_M+\frac1p\right).
\]
Its derivative,
\[
 L'(p)
 =
 \log\frac{\alpha}{\beta_M}
 -
 \frac{\alpha+\beta_M}
 {[p(1-\alpha)+1][p(1+\beta_M)+1]},
\]
is strictly increasing in \(p\).  Thus \(L\) is strictly convex.  In addition,
\[
 c_0(\alpha_H)
 =
 -\frac{\beta_M^4}{(1-\beta_M)^2}<0,
\]
and \(c_0<0\) for every \(\alpha\leq\alpha_H\), while
\(L(p)\to+\infty\) when \(\alpha>\beta_M\).  The coefficient sequence can
therefore change sign at most once, and only from negative to positive.

\begin{lemma}[Power-series one crossing]\label{app:lem:series-crossing}
If a power series on \(t\in(0,1)\) has coefficients that change sign at most
once, from negative to positive, then the series has at most one zero and can
cross only from negative to positive.
\end{lemma}

\begin{proof}
Let \(K\) be the first index with a nonnegative coefficient and suppose the
series is zero at \(t_0\).  For \(t>t_0\), every term in the positive tail is
amplified by at least \((t/t_0)^K\), whereas every negative-head magnitude is
amplified by at most \((t/t_0)^{K-1}\).  The new sum is strictly positive.
\end{proof}

Applying the lemma to \(t=y^2\) in \eqref{app:eq:F-series} shows that
\(F_{M,y}\) starts negative and changes sign at most once.  At
\(\alpha=\alpha_H\), the equal endpoint values then imply
\begin{equation}\label{app:eq:F-H-negative}
 F_M(y,\alpha_H)<0,\qquad 0<y<1.
\end{equation}
Together with \eqref{app:eq:F-alpha} and the value at \(\beta_M\), this gives
a unique root
\[
 \alpha_\infty(y)\in(\beta_M,\alpha_H),
 \qquad 0<y<1.
\]
At \(y=0\), the endpoint root is
\(\alpha_\infty(0)=\alpha_H\).  At an interior root, \(F_{M,y}<0\): if the
single derivative crossing exists,
an interior level-zero crossing must occur on the decreasing branch because
the increasing branch ends at \(F_M(1,\alpha)=0\).

\begin{theorem}[High-curvature pair ordering]
\label{app:thm:high-curvature-order}
The limiting threshold is strictly increasing in the smaller residual share
on \(x\in(0,1/2]\).  Its derivative satisfies
\[
 \frac{d\delta_\infty^B(x)}{dx}>0,
 \qquad
 \frac{dq_\infty^B(x;n)}{dx}>0,
 \qquad 0<x<\frac12,
\]
and both derivatives equal zero at \(x=1/2\).
Consequently, for fixed integer \(n\), among the feasible grid
\[
 \mathcal X_n
 =
 \left\{\frac1n,\frac2n,\ldots,
 \frac{\lfloor n/2\rfloor}{n}\right\},
\]
the canonical share \(x=1/n\) is the unique earliest two-source block in the
high-curvature limit.
\end{theorem}

\begin{proof}
For \(0<y<1\), along the root,
\[
 \alpha_\infty'(y)
 =-\frac{F_{M,y}}{F_{M,\alpha}}<0.
\]
Because \(y=1-2x\), the root \(\alpha_\infty\) rises with \(x\).  Both
\(\delta=2\alpha/(1+\alpha)\) and
\(q=\delta/[n-(n-1)\delta]\) rise with \(\alpha\), giving the strict interior
derivatives.  At \(x=1/2\), \(y=0\); evenness of \(F_M\) in \(y\) gives
\(F_{M,y}=0\), so both endpoint derivatives are zero.  Continuity together
with the strict interior ordering makes both threshold functions strictly
increasing on the half-closed domain.
\end{proof}

\subsection{Returning to finite curvature}

The expansion \eqref{app:eq:S-expansion} is uniform in a neighborhood of each
simple root, and both the finite and limiting pair sums strictly decrease in
\(\delta\).  Standard monotone root convergence gives
\[
 \delta_r^B(x)\longrightarrow\delta_\infty^B(x).
\]
For a fixed integer \(n\), the feasible set
 \(\mathcal X_n\) displayed in
Theorem~\ref{app:thm:high-curvature-order}
is finite.  Theorem~\ref{app:thm:high-curvature-order} leaves a strictly
positive minimum gap between the limiting canonical threshold and every other
grid threshold.  Root convergence therefore preserves the ordering beyond a
finite curvature.

\begin{corollary}[Eventual canonical-first]
\label{app:cor:eventual-canonical}
For every fixed \(M,n\geq2\), there is a finite
\(\bar\rho(M,n)\) such that, on the matched locus,
\[
 \rho\geq\bar\rho(M,n)
 \quad\Longrightarrow\quad
 q_1^B(\rho)<q_t^B(\rho)
 \quad
 \text{for }t=2,\ldots,\lfloor n/2\rfloor.
\]
Combined with Theorem~\ref{app:thm:two-source} and full-size completion, the
canonical pair then determines the exact complete one-new-team stability
threshold.
\end{corollary}

The limiting result also explains numerical near ties.  Let \(x_0=1/n\) and
take any noncanonical \(x>x_0\).  At the finite canonical threshold,
\[
 S_r\!\left(x,\delta_r^B(x_0)\right)
 =
 \frac{J_{\delta_\infty^B(x_0)}(x)}{r}
 +O(r^{-2})>0.
\]
The noncanonical pair remains unprofitable, but its probability slack is only
of order \(1/(\rho-1)\).  The asymptotic convergence is therefore slow enough
to require high precision in large-curvature evaluations.

\end{document}